\documentclass[a4paper,11pt]{article}
\pdfoutput=1 
\usepackage{jcappub}
\usepackage{amsmath}
\usepackage{amssymb}
\usepackage{rotating}
\usepackage[T1]{fontenc} 

\def\({\left(}
\def\){\right)}
\def\[{\left[}
\def\]{\right]}
\def\be{\begin{equation}}
\def\ee{\end{equation}}
\def\beq{\begin{eqnarray}}
\def\eeq{\end{eqnarray}}

\title{\boldmath A novel probe of ionized bubble shape and size statistics of the epoch of reionization using the contour Minkowski Tensor}

\author[a,b]{Akanksha Kapahtia} 
\author[a,1]{Pravabati Chingangbam,\note{Corresponding author}}
\author[c]{Stephen Appleby} 
\author[c]{Changbom Park}

\affiliation[a]{Indian Institute of Astrophysics, Koramangala II Block,
  Bangalore  560 034, India}
\affiliation[b]{Joint Astronomy Program, Department of Physics, Indian Institute of Science, C. V. Raman Ave., Bangalore  560 012, India} 
\affiliation[c]{Korea Institute for Advanced Study, 85 Hoegiro, Dongdaemun-gu,  
  Seoul 02455, Korea}

\emailAdd{akanksha.kapahtia@iiap.res.in}
\emailAdd{prava@iiap.res.in}
\abstract{
  We employ the rank-2 {\em contour} Minkowski tensor in two dimensions to probe length and time scales of ionized bubbles during the epoch of reionization. We demonstrate that ionized bubbles are  not circular, and hence not spherical in three dimensions, as is often assumed for simplified analytic arguments. We quantify their shape anisotropy by using the ratio of the two eigenvalues of the contour Minkowski tensor.  Further, we show that a size parameter constructed from the eigenvalues of this tensor provide an upper bound on the characteristic sizes of ionized bubbles at different redshifts. The shape and size parameters provide information of the characteristic time epochs when bubble mergers begin and end. Our method is expected to be useful to reconstruct the reionization history using data of the brightness temperature field. }
   
\begin{document}
\maketitle
\flushbottom
\section{Introduction} 

After the epoch of recombination most of the baryonic matter in the Universe must have been in the form of neutral hydrogen. The neutral hydrogen is expected to emit 21-cm radiation from the transition between the hyperfine states of the ground state of neutral hydrogen.   The detection of this radiation  from the Epoch of Reionization (EoR)~\cite{Madau:1997} is one of the most exciting frontier prospects of observational cosmology. This signal will be observed in frequencies that are suitably redshifted from its rest frame value $\sim$1420 MHz. Since the EoR ends at redshift 6, the 21-cm signal will be observed at frequencies lower than 203 MHz. The spatial distribution of the 21-cm emission will contain imprints of reionization physics (see e.g. \cite{Furlanetto:2006jb}), as well as the nature and growth of the primordial density perturbations. The EoR commences when the first ionizing sources get formed, and progresses as more sources get formed and they ionize the neutral hydrogen surrounding them.

The size of ionized regions (or bubbles) and their spatial distribution is an important observable for probing the EoR. The growth of their sizes with redshift, and distribution, have been addressed by several authors~\cite{Cen:2005ax,Wyithe:2004qd,Furlanetto:2004nh,Furlanetto:2005ax,McQuinn:2006et,Zahn:2006sg,Cohn:2006ge,Feng:2012mab,Aseem:2014,Lin:2015bcw,Giri:2017nty}, using simulations and analytic approaches. It has been reported that ionized regions can grow to several tens of Mpc. In simplified analytic calculations, ionized bubbles are usually approximated to be spherical. The ionized region close to the ionizing source may very well be spherical following the shape of the source. However, the propagation of the ionized front in different directions away from the source depends on factors such as matter density fluctuations, recombination of hydrogen, clumping of the inter-Galactic medium, etc. Hence, it is unlikely that the ionized bubbles will maintain isotropy. In this paper we address the quantification of the characteristic shape of ionized regions as well as the growth of their sizes from a new unified perspective. Further, we relate the redshift evolution of the shape and size of ionized regions to the progress of reionization. Earlier discussions of the shape of ionization fronts near high redshift quazars can be found in~\cite{Feng:2012mab,Majumdar:2011uy}.
 
The detection of 21-cm signal is experimentally extremely challenging. Apart from instrumental noise,  the signal is dominated by foreground emissions by several orders of magnitude. Therefore, a thorough understanding of the 21-cm signal, instrumental noise and characteristics of foregrounds emissions is crucial for interpreting observed data and will guide towards statistical detection of the signal (see e.g.~\cite{Zaroubi}). There has been a claim for the detection of the global 21-cm signal recently~\cite{Bowman:2018} and confirmation from other ongoing experiments is awaited.  
The spatial distribution of the 21-cm signal, when detected, will help tighten cosmological constraints, complementary to the CMB and large scale structure. 
It is hoped that upcoming radio interferometer experiments such as the Square Kilometer Array (SKA)~\cite{Carilli:2015yta} will be able to detect the signal. 

To unlock the physical information encoded in any cosmological field it is very important to devise efficient statistical tools. The two-point function has been popularly used for this purpose. However, the fields of the EoR are expected to show strong departure from Gaussian nature and hence the two-point function is of limited use. The three-point function has been employed~\cite{Ali:2005md,Yoshiura:2014ria,Shimabukuro:2015iqa,Shimabukuro:2016viy,Majumdar:2017tdm} to capture some non-Gaussian nature of the fields. Given the expected strong non-Gaussian nature of the fields it is desirable to use statistical methods which can capture all orders of $n$-point functions. 
In this paper we use the rank-2 {\em contour Minkowski Tensor} (CMT), which is one of the Minkowski Tensors~\cite{McMullen:1997,Alesker:1999, Beisbart:2002,Hug:2008,Schroder2D:2009,Schroder3D:2013} that were recently introduced for cosmological random fields in two dimensions~\cite{Vidhya:2016,Chingangbam:2017,Appleby:2017uvb}. Minkowski Tensors are tensor generalizations of the scalar Minkowski Functionals (MFs)~\cite{Tomita:1986,Gott:1990,Mecke:1994,Schmalzing:1997,Schmalzing:1998,Winitzki:1998,Matsubara:2003yt}. In particular, the CMT is the tensor generalization of the second scalar MF, the total contour length. 
Scalar MFs have been used in many areas of cosmology (see e.g. ~\cite{COBE_NG:2000,Park:2002,Komatsu:2011, Chingangbam:2012wp,Vidhya:2014,Ade:2015ava,Buchert:2017uup}).   In the context of 21-cm emissions, the genus, which is one of the scalar MFs, has been used to study different reionization models and history~\cite{Lee:2007dt,Friedrich:2010nq,Ahn:2010hg,Wang:2015dna}. MFs in three dimensions have also been used for similar purpose~\cite{Gleser:2006su,Yoshiura:2015}.

The CMT carries information of the anisotropy and relative alignment of structures in two dimensions. In this paper, we use the CMT as a tool to address
two important issues concerning the EoR, namely, quantifying the shapes of ionized bubbles, and finding their characteristic size~\cite{Furlanetto:2004nh,McQuinn:2006et,Aseem:2014,Lin:2015bcw,Giri:2017nty} at different redshifts.  
We apply the CMT to  simulated fields of the EoR made using \texttt{21cmFAST}~\cite{Zahn:2010yw,Mesinger:2010ne}.
The eigenvalues provide information of the size, and their ratio quantifies the shape anisotropy, of the ionized regions. We show that statistically the typical shape of ionized regions departs from isotropy, and the evolution of the level of anisotropy is determined by the reionization history. We demonstrate how the characteristic bubble size and shape, and their number counts, are related to the characteristic time epochs when fragmentation of neutral regions occur, counts of neutral regions and ionized bubbles become equal, and bubble mergers begin and end. 
Our paper thus presents a proof-of-concept of how the reionization history is encoded in the  characteristic shape and size of the ionized and neutral hydrogen regions. We focus mainly on the ionization field, and its effect on the differential brightness temperature. A full analysis of all the fields of the EoR and comparison of different models of reionization will be pursued in our future work.

It is important to note that our analysis is carried out using idealized simulations. We have not considered the impact of foreground contamination and instrument noise on our results. Given that the signals of the frequencies of interest are dominated by foreground signals by several orders of magnitude we can foresee that a sound understanding of the properties of the foregrounds themselves, as well as instrumental effects, will be necessary in order to reliably use our method on real data.  
 Currently the best constraint on reionization history is obtained using data of the Cosmic Microwave Background (CMB) radiation from PLANCK~\cite{PLANCKreio:2016,PLANCK:2018_cosmoparams}, which gives the integrated optical depth during the EoR. The precise details of the results that we present here will depend on the reionization history and hence the constraint from PLANCK will be important for discriminating models. However, the general conclusions that we draw in this first paper will remain valid. 

The following sections are organized as follows. Section \ref{sec:sim} briefly describes the physics of the EoR. Section \ref{sec:tmf} reviews the definition of the CMT and the size and shape information encoded in it. In section \ref{sec:tmf_reio} we present out results for the characteristic sizes and shapes of the ionization field and the differential brigthness temperature. We end with a summary and discussion of our results and future directions in section \ref{sec:conclusion}.

\section{21-cm Brightness Temperature Field}
\label{sec:sim}

The energy difference between the hyperfine levels of the neutral Hydrogen in ground state corresponds to an excitation temperature of $T_*= 0.068$~K. The emission or absorption for this transition is determined by the spin temperature, $T_s$, which is the temperature at which the relative population of the two levels become $n_{1}/n_{0}=3~$exp~$(-T_{*}/T_{s})$, if the system is in equilibrium. 
Since the redshifted frequency of this transition lies in the radio regime, the intensity of this spectral line is quantified by the brightness temperature $T_b$. The brightness temperature is measured in emission or absorption against a background of CMB. From the equation of radiative transfer along the line of sight, the offset of the 21-cm brightness temperature from the CMB temperature, usually referred to as the {\em differential brightness temperature}, $\delta T_{b}$, is given by:
\begin{eqnarray}
\delta T_{b}=\frac{T_{s}-T_{\gamma}}{1+z}(1-e^{- \tau}),
\end{eqnarray}
where $\tau_{\nu}$ is the optical depth for the 21-cm emission integrated along the line of sight up till the redshift of observation, $z$, and $T_{\gamma}$ is the CMB temperature.  The differential brightness temperature for an observed frequency $\nu$ is then given by the following expression~\cite{Furlanetto:2006jb}:
\begin{equation}
  \delta T_b(\nu) \approx 27\,x_{\rm HI}\left(1+\delta_{\rm NL}\right)\bigg(\frac{H}{dv_r/dr+H}\bigg)  \bigg( 1-\frac{T_\gamma}{T_S}\bigg) \,\sqrt{\frac{1+z}{10}\frac{0.15}{\Omega_Mh^{2}}}\,\frac{\Omega_bh^2}{0.023} \,{\rm mK},
\label{eqn:dTb}
\end{equation}
where $x_{\rm HI}$ is the neutral hydrogen fraction, $\delta_{\rm NL}$ is the density contrast due to non linear density evolution up till the redshift of interest,  $dv_r/dr$ is the derivative of the peculiar velocity of the hydrogen gas along the line of sight. $\Omega_b$ and $\Omega_M$ are the usual matter density fractions for baryons and dark matter, respectively. $H$ is the Hubble parameter at redshift $z$ and $h$ is the dimensionless Hubble parameter today.   

In order to study the ionization history, we have generated mock 21-cm fields in a 200 Mpc box with periodic boundary conditions, using the publicly available semi numerical code \texttt{21cmFAST v1.2}~\cite{Mesinger:2010ne}. The code generates Gaussian random initial density fields and then evolves it using the Zel'dovich approximation. It generates the density $~\delta_{\rm NL}(\vec x)$, $~T_{s}(\vec x)$ , peculiar velocity and $x_{\rm HI}(\vec x)$ at every grid point on a regular lattice and finally calculates $\delta T_b (\vec x)$ at these points. For our purpose we choose $1024^3$ pixel grid for the initial conditions and $512^3$ pixel grid for the evolved fields. The initial conditions were generated at a redshift of $z=300$. The background cosmology we have adopted is given by the PLANCK 2015 flat $\Lambda$CDM parameter values~\cite{planck:cosmopara2015} given by $\Omega_M=0.308,~\Omega_b=0.02226/h^2,~n_s=0.968,~\sigma_8=0.815, ~h=0.678$, where $n_s$ is the spectral index of primordial fluctuations and $\sigma_8$ is the amplitude of matter fluctuations at $8\, h^{-1}$Mpc.

In order to identify ionized regions the code uses an excursion set approach similar to the Press-Schechter theory of halo mass functions. According to this approach \cite{Furlanetto:2004nh} an isolated region of mass $m$ is self ionized if it has sufficient mass in luminous sources. For such a region of size $R(m)$ corresponding to the mass $m$ the number of ionizing photons should be greater than the number of neutral hydrogen atoms. The above criteria boils down to the following condition, 
\begin{equation}
f_{\rm coll} (\vec x,z,R)\geq\zeta^{-1},
\end{equation}
where $f_{\rm coll}$ is the fraction of mass residing in collapsed halos inside a sphere of mass $m=4/3 ~\pi R^{3}~ \overline{\rho}~ [1+\langle \delta_{\rm NL} \rangle_{R}]$. $\zeta$ is a factor that describes the amount of mass a galaxy of mass $m_{\rm gal}$ can ionize: $m_{\rm ion}=\zeta m_{\rm gal}$. $\zeta$ encapsulates information about the efficiency of ionization of the luminous sources during EoR. Therefore different values of $\zeta$ describe different ionization histories. For our calculations here we have chosen $\zeta=50$.

In \texttt{21cmFAST} a central cell is flagged as ionized if the aforementioned condition is fulfilled at some filter scale while reducing from a maximum value $R_{\rm max}$ to cell size in logarithmic steps. We use $R_{\rm max}=10~$ Mpc. IGM heating is considered throughout the evolution (i.e. $T_s$ is not ignored at lower $z$ values). The X-ray efficiency parameter is chosen to be $\zeta_x=2 \times 10^{56}$ and X-ray spectral index of 1.2. The minimum virial temperature of haloes is chosen to be $T_{vir}=3\times 10^4 ~K$.

A choice of values of the various physical parameters described above typically leads to a particular reionization history, but there can be degeneracies between the parameters. In this paper we have chosen the parameter values described above that correspond to a realistic history of reionization, such as given by constraints from the 2015 PLANCK data~\cite{PLANCKreio:2016}. Further, we assume that peculiar velocity is much smaller than the Hubble expansion rate. Therefore, we drop the factor containing the peculiar velocity in Eq. (2.2), and postpone analysis of its effect to future work.

\subsection{Progress of reionization and length and time scales}

\begin{figure}
\begin{center}
	\resizebox{5.5in}{5in}{\includegraphics{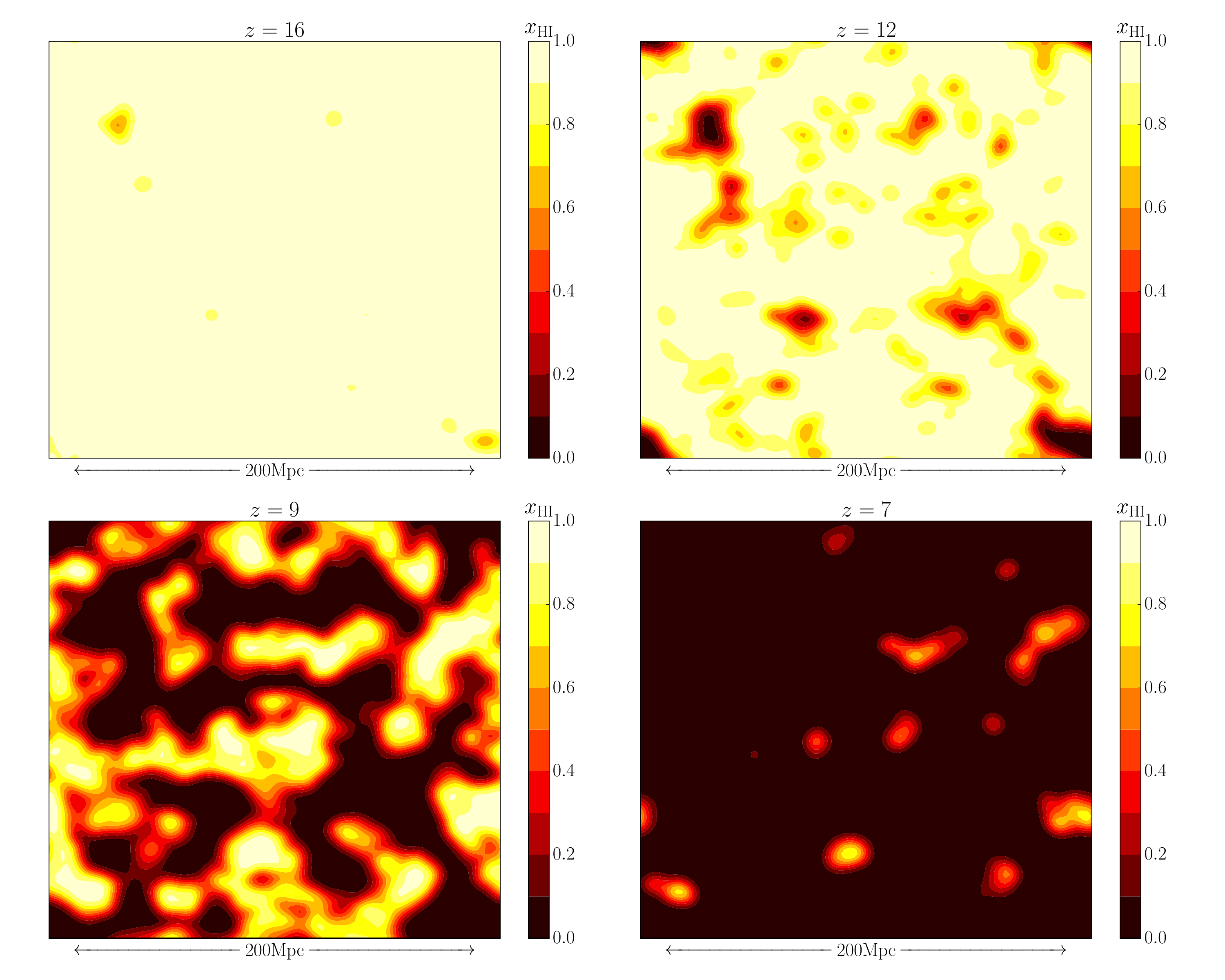}} 
	\end{center}
	\caption{The four images show a two dimensional slice of $~x_{\rm HI}(\vec x)$ at four different redshifts for the model of interest here. These images highlight the redshift evolution of the numbers, shapes and sizes of the connected regions (pale yellow) corresponding to neutral regions, and hole regions (varying from dark yellow to black) corresponding to ionized regions. }
	\label{fig:xh_slice}
\end{figure}

The focus in this paper is to track the redshift evolution of the numbers, sizes and shapes of neutral (connected) regions and ionized regions and determine their statistics. We use the term {\em ionized bubble} to refer to only those ionized regions that are enclosed by neutral regions. They are also referred to as {\em holes}. Before we carry out quantitative analysis in section 4 it is instructive to use physical reasoning and anticipate their redshift evolution. 

In Fig~(\ref{fig:xh_slice}) we show the progress of reionization for a two dimensional slice of $~x_{\rm HI}(\vec x)$ at four different redshifts. In all the panels the pale yellow regions are the neutral regions while the red are the ionized regions.  At the relatively high redshift $z=16$ the ionized regions are all bubbles. They are small, isolated and tend to be circular (though not exactly circular). During this early period of the EoR the number of ionized bubbles increases with time because new ionizing sources are getting formed, and the size of the ionized regions grow. 
As the bubble sizes grow, two or more bubbles can merge. As bubble mergers take place the shape of the bubbles will become more anisotropic. Therefore, the average number, size and shape of ionized regions at any redshift will be governed by three factors: (1) the rate of formation of ionizing sources, (2) the rate at which the ionized regions grow, and (3) the rate of bubble mergers. These factors will in turn depend on physical parameters such as $\zeta$ and $T_{\rm vir}$. 
We expect that at early redshifts the rate of formation of ionizing sources will be greater than the rate of bubble mergers, and vice versa at later redshifts. We refer to the redshift at which they are equal as $z_{\rm frag}$. Around this redshift we also expect that the single connected (neutral hydrogen) region will begin to fragment into multiple smaller connected regions. Therefore, at $z_{\rm frag}$ the number of ionized bubbles will turn over and start decreasing, while the number of connected regions will start increasing. The anisotropy of the shapes of the ionized bubbles within the neutral regions will also show an increase.  

Let $z_{0.5}$ denote the redshift when $\bar{x}_{\rm HI}=0.5$. As reionization progresses there must be an epoch when the number of connected regions and holes cross each other and become equal. We will show in section 4 that the equality happens at $z_{0.5}$. Then, even further as $z$ decreases beyond $z_{0.5}$, the merged bubbles will still grow in size with a corresponding decrease of anisotropy. At the same time, the number of connected regions will increase as more and more fragmentation occurs, with a corresponding decrease in size. This process will continue up to a redshift $z=z_{\rm e}$, after which the number of connected regions will drop due to ionization of the entire region reaching completion and the number of holes will go to zero. 

We will determine $z_{\rm frag}$, $z_{0.5}$ and $z_{\rm e}$ by using the contour Minkowski Tensor in section 4.  We will then quantify the anticipated evolution described above for the numbers of ionized bubbles and connected regions, their sizes and anisotropy.

\section{Definition of Contour Minkowski Tensor}
\label{sec:tmf}

Minkowski Tensors (MTs) are tensor generalizations of the scalar MFs. For our analysis the kind of Euclidean two-dimensional structures we are interested in are connected regions such as discs, and holes within connected regions. MTs can be defined for the boundary curves of these structures. We restrict attention to the following MT, which we refer to as the {\em contour} MT, defined for a single boundary curve, $C$,  as, 
\begin{equation}
\mathcal{W}_{1}= \int_C \hat{T}\otimes \hat{T} ~{\rm d}s,
\label{eqn:W1}
\end{equation}
where $\hat{T}$ is the unit tangent vector at every point on the curve, $\otimes$ denotes the symmetric tensor product given by
\begin{equation}
\left(\hat{T}\otimes \hat{T}\right)_{ij} = \frac12\left( \hat{T}_i\hat{T}_j + \hat{T}_j\hat{T}_i\right),
\end{equation}
and ${\rm d}s$ is the arc length. Our notation follows~\cite{Chingangbam:2017}. $\mathcal{W}_{1}$ is referred to as ${W}_2^{1,1}$ in \cite{Schroder2D:2009,Vidhya:2016,Appleby:2017uvb}. However, note that the numerical factor on the r.h.s. of Eq. \ref{eqn:W1} is different from these earlier papers, and it has been chosen so that 
\begin{equation}
{\mathbf{Tr}}\left(\mathcal{W}_{1}\right) = \int_C ~{\rm d}s,
\label{eqn:W1_perimeter}
\end{equation}
which is simply the second scalar MF, the total contour length or perimeter of the curve denoted by $W_1$. 

$\mathcal{W}_{1}$ is translation invariant and transforms as a rank-2 tensor under rotations. It has dimension of length. Since it is symmetric and defined for a closed curve, its two eigenvalues denoted by $\lambda_1$ and $\lambda_2$ are real and positive. For convenience of notation we choose  $\lambda_1<\lambda_2$. When the eigenvalues are different they pick out two orthogonal directions and we can effectively approximate the arbitrary curve as an ellipse whose semi-minor axis is aligned with the eigenvector of $\lambda_1$ while the semi-major is aligned with that of $\lambda_2$. 

\begin{figure*}
 \begin{center}
   \resizebox{3.1in}{2.2in}{\includegraphics{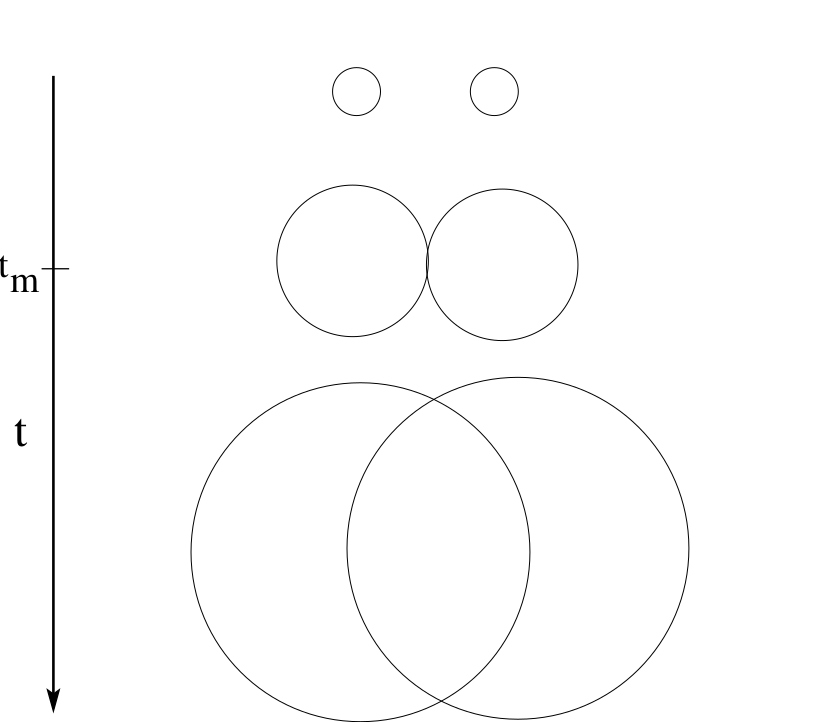}}\quad\quad\quad
   \resizebox{2.4in}{2.2in}{\includegraphics{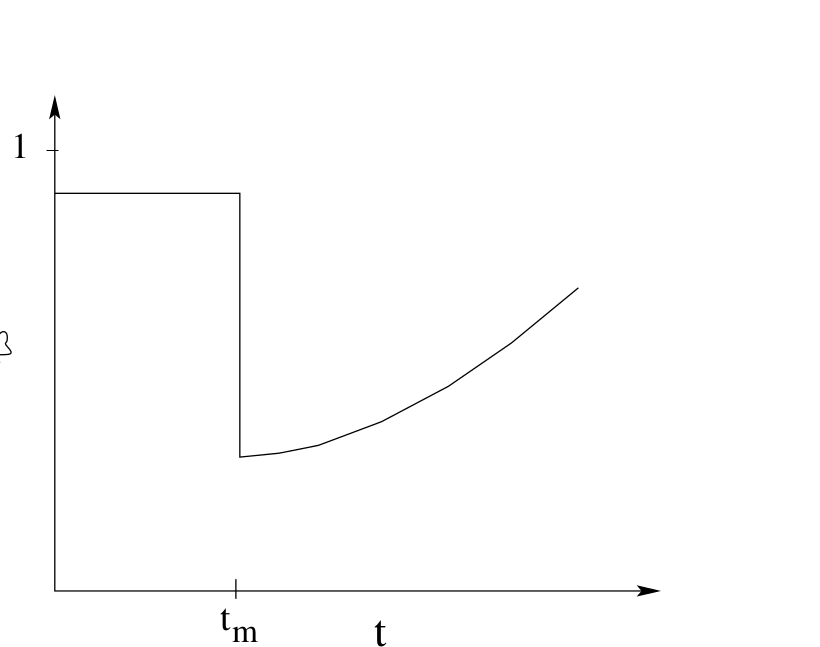}}   
 \end{center}
 \caption{{\em Left:} A simplistic depiction of two elliptic regions that grow, merge to become a single region at time $t=t_{\rm m}$, and then grow further (top to bottom). {\em Right}: The average value of $\beta$ will be constant till the two ellipses merge at $t_{\rm m}$, at which time $\beta$ will drop sharply, and then increase again as the combined region grows further. This depiction captures the essence of the time evolution of the shapes of ionized regions that grow, merge and grow further.}
 \label{fig:bubblemerger}
\end{figure*}

{\em Isotropy of closed curves}:  The ratio $\beta\equiv\lambda_1/\lambda_2$ is called the {\em shape anisotropy parameter} of the curve. We define a curve to be isotropic if $\beta=1$. Isotropic curves are those that  have $m$-fold, $m\ge 3$, rotational symmetry. Note that circular shape is just one of the possible isotropic shapes, given by the limit $m\rightarrow\infty$. For a generic curve, $\beta$ will have value between zero and one. 
Let us consider a toy example of how $\beta$ can evolve with time. The left panel of Fig. \ref{fig:bubblemerger} shows a simplistic example of two ellipses of identical shape created far away from each other at some initial time (top, left).  Shown on the right is the evolution of $\beta$. At the time the ellipses are still apart the value of $\beta$ will be some value lower than one. As the ellipses grow in time, $\beta$ remains the same, till they merge to form one single region (left, middle) at time $t=t_{\rm m}$. At this time $\beta$ will drop since the resultant shape is highly anisotropic. Then, as the regions grow further, $\beta$ will again start rising (left, bottom). This simplified scenario will be relevant when we apply the CMT to the EoR.

The CMT also gives an estimate of the size of the area enclosed by the curve. Let $\lambda\equiv \lambda_1+ \lambda_2$ denote the perimeter of the closed curve. Then, using this perimeter to define a circle by equating  $\lambda$ to $2\pi r$, we determine $r$ to be
\begin{equation}
  r\equiv\lambda/2\pi.
\end{equation}
For an arbitrary closed curve the isoperimetric inequality relates the area, $A$, enclosed by the curve, to its perimeter $\lambda$ as
\begin{equation}
  A \le \frac{\lambda^2}{4\pi}. 
\end{equation}
This implies that $r$ will in general result in an overestimation of the size of the area enclosed by the curve. The overestimation will be more for non-convex curves\footnote{A closed curve in Euclidean space is called {\em convex} if the straight line joining any two arbitrary points in the region enclosed by the curve also lies within the same region. It is called {\em non-convex}  if there is at least one pair of points such that some part of the line joining them lies outside the region.}. Therefore, $r$ actually gives an upper bound of the size.  We emphasize that the perimeter of the structure can be calculated exactly, {\em it is only the interpretation of the idealized circle that $r$ over-estimates.} 
For determining the radius for a single structure knowledge of $W_1$ is sufficient  and we do not require the CMT.  However, when we have an ensemble of structures, in addition to $W_1$, the number of curves (the so-called Betti numbers, see section 4), is required  to obtain the estimate of the mean size of structures.  The truly new information that we obtain from the CMT is that of the shape of the curve. Nevertheless, it neatly unifies the size and shape information in a single tensor quantity, and we employ it to obtain estimates of the size of structures in conjunction with the Betti numbers.

Eq. \ref{eqn:W1} defines $\mathcal{W}_{1}$ for a single curve. We next shift attention to smooth random fields. At each  given threshold value of a random field the boundaries of the excursion or level set enclose either connected regions or holes. We therefore obtain sets  of these two types of curves for every threshold. The morphological properties of the excursion sets, and as a consequence that of these boundaries, change smoothly as the threshold value is varied. We can then compute 
$\mathcal W_1$ for each of the boundary curves at a finite set of threshold values. 

It is also useful to express $ {\mathcal W}_1$ in terms of derivatives of the field. Let $u$ denote the field. Then in terms of its derivatives $u_i$ the components of the unit tangent vector $\hat T$ can then be written as,
 \begin{equation}
  \hat{T}_i = \epsilon_{ij} \, \frac{u_{j}}{\left| \nabla u \right|},
  \label{eqn:hatt}
 \end{equation}
 where $\epsilon_{ij}$ is the anti-symmetric Levi-Civita tensor in two dimensions. 
 $\mathcal{W}_1$ for a particular curve $C$ can then be expressed as,
 \begin{equation}
  {\mathcal W}_1 = \int_{C}  {\rm d}\ell  \ \frac{1}{|\nabla u|^2} \ {\mathcal M},
  \label{eqn:W1u}
 \end{equation}
 where the matrix $\mathcal{M}$ to be evaluated at the points along the curve is given by,
 \begin{equation}
  \mathcal M=  \left(
  \begin{array}{cc} 
    u_{2}^2 &  u_{1} \,u_{2} \\
      u_{1} \,u_{2} & u_{1}^2
  \end{array}\right).
  \label{eqn:M}
 \end{equation}
 It is again easy to see in this form that the trace of  $\mathcal W_1$ gives $W_1$.
 
 For any field, subtracting its mean values merely shifts the field values and does not change its geometrical and topological properties. A rescaling $u \rightarrow \tilde u\equiv u/a$, where $a$ is some constant, at every spatial point also does not alter the topology and the geometry of the excursion sets even though field values get remapped. This is obvious from the r.h.s. of Eq. \ref{eqn:W1u} and \ref{eqn:M} where the integrand is clearly independent of field rescalings by constant factors.

 Our numerical calculation follows the method described in~\cite{Appleby:2017uvb}. It involves finding the boundary contours for each threshold, identifying each closed curve bounding either a connected region or a hole, and then calculating $\mathcal W_1$ for each curve. 
 In this work the fields are simulated with periodic boundary conditions and hence we do not need to consider the effect of boundary cuts. However, when applying the statistics to actual data, the brightness temperature field will only be measured over a limited region of the sky. 
 Connected regions and holes that coincide with survey boundaries should be disregarded when taking the average $\beta$, and the effect of the mask must be assessed by applying it to Gaussian random fields and also simulated fields. 

Note that we could calculate the area of each structure directly and also roughly estimate its shape anisotropy by finding its centre of mass and taking the ratio of the smallest and largest distances of the boundary from the centre. The calculational advantage of the CMT over such ad hoc methods becomes clear when we apply to a large collection of curves, such as those given by the boundaries of excursion sets. The CMT is superior to simplistic approaches because they are based on very general mathematical framework with well defined transformation properties. This makes them unambiguous and easy to apply to any distribution of structures.

\section{Contour Minkowski Tensor for reionization fields}
\label{sec:tmf_reio}

\begin{figure*}
  \begin{center}
    \includegraphics[width=4.5cm, height=7.5cm,angle=-90]{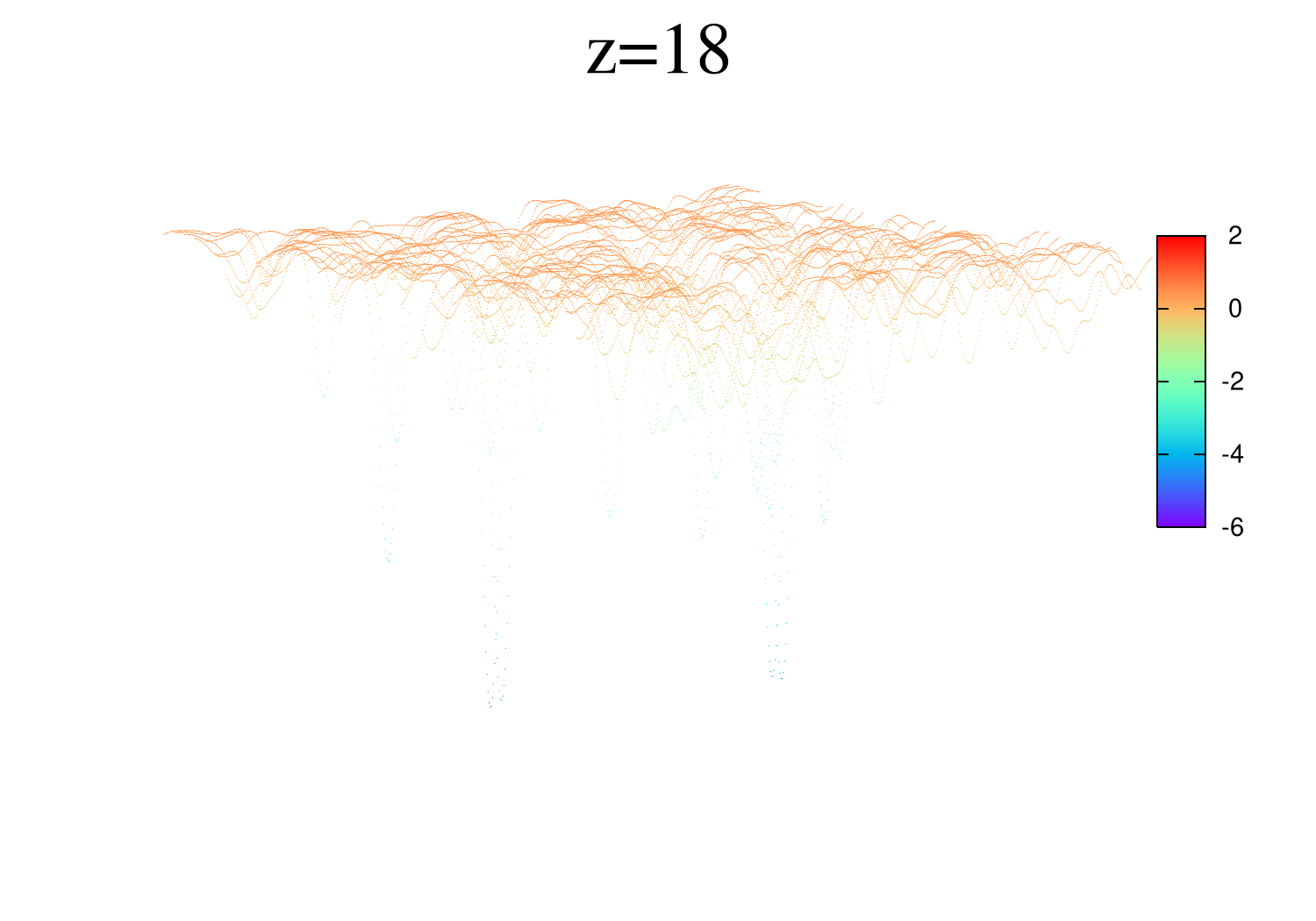}\quad
    \includegraphics[width=4.5cm, height=7.5cm,angle=-90]{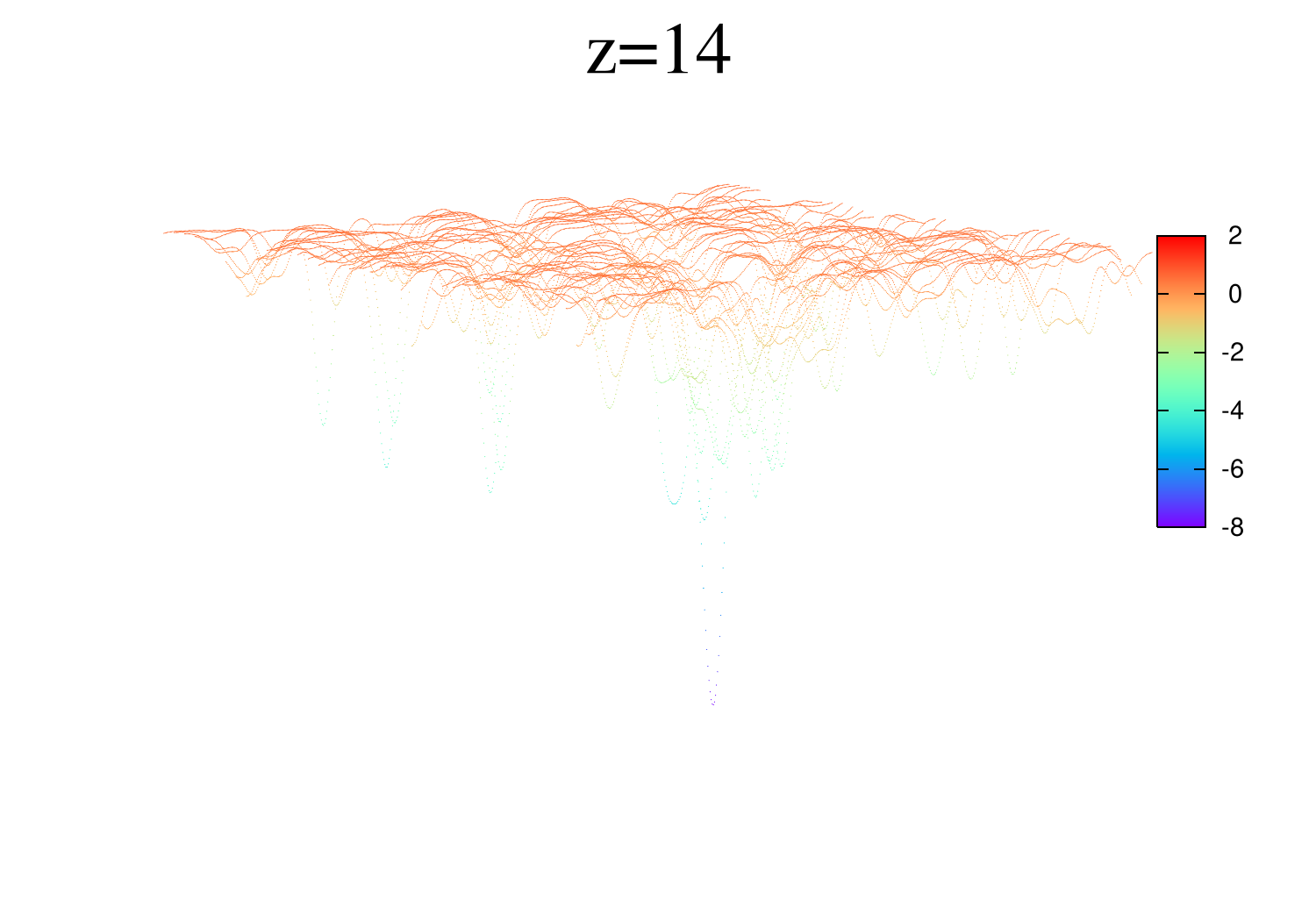}
    \includegraphics[width=4.5cm, height=7.5cm,angle=-90]{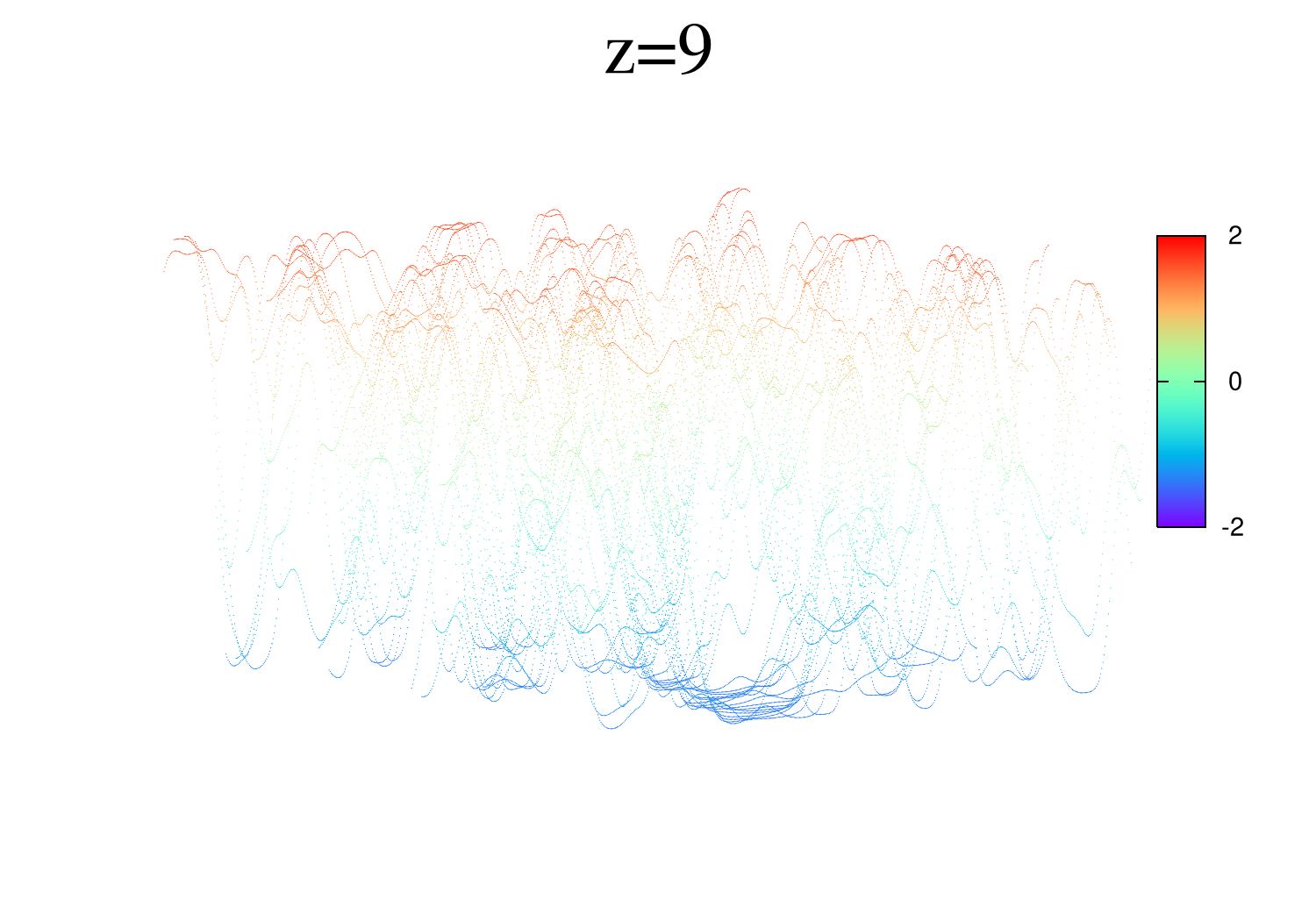}\quad
    \includegraphics[width=4.5cm, height=7.5cm,angle=-90]{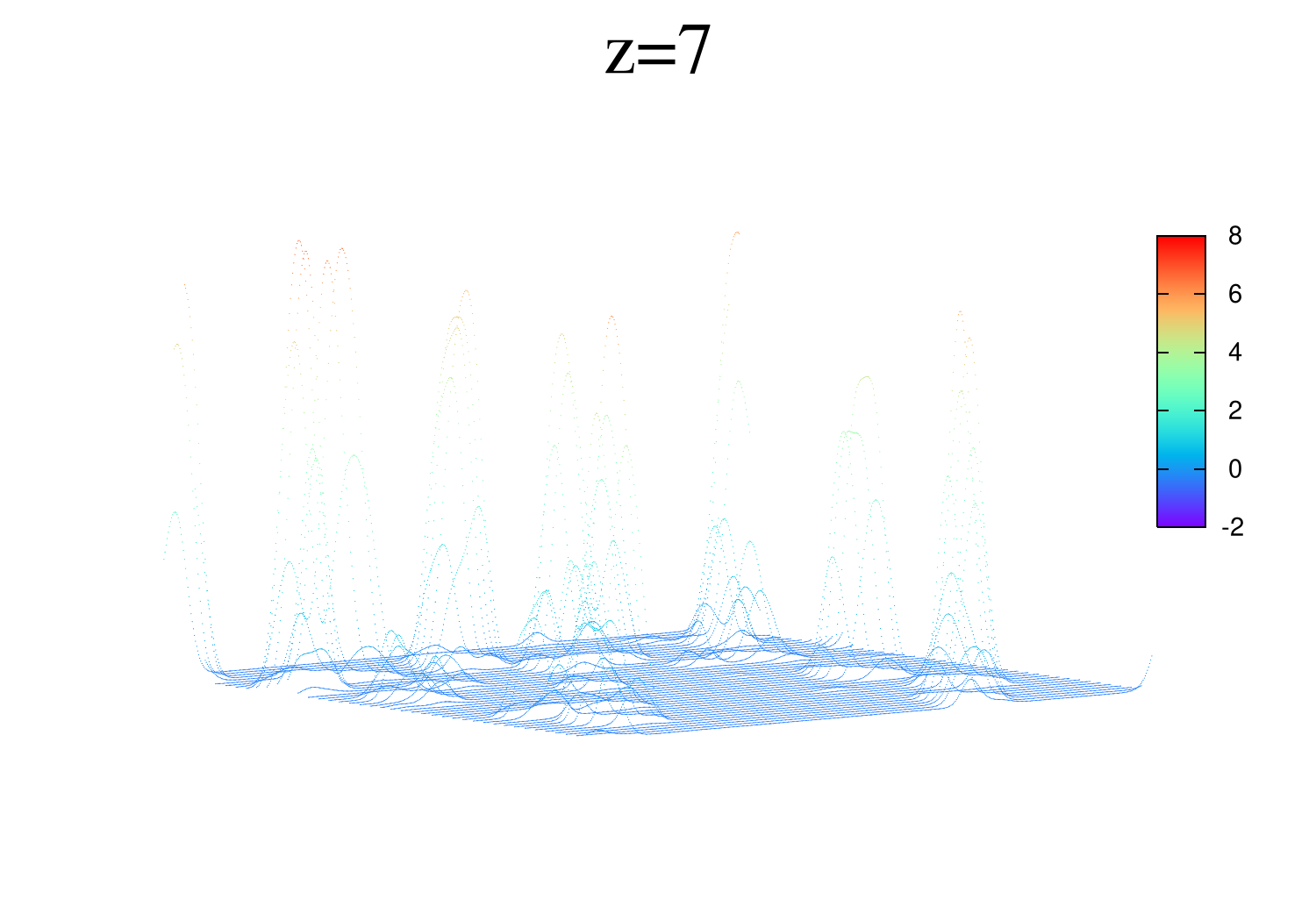}
  \end{center}
  \caption{Redshift evolution of a slice of the $x_{\rm HI}$ field. These figures carry the same information as Fig.~\ref{fig:xh_slice}, but presented in a form that makes the visualization of the field levels easy.}  
  \label{fig:xH_slice}
\end{figure*}
We now focus on the fields of the EoR, namely, 
$x_{\rm HI}$ and $\delta T_b$. We generate them  at several redshift values and  smooth  them with Gaussian smoothing kernel using different values of the smoothing scale, $R_s$. Each of these fields, say $u$, is redefined to the standard normal form, $u\rightarrow \tilde{u} \equiv (u-\mu)/\sigma$, where $\mu$ is the mean and $\sigma$ is the standard deviation of $u$. As discussed in section 3, this redefinition does not alter its geometrical and topological properties, but allows for uniform choice of threshold values for different fields. 
Then, we divide the field into two dimensional slices. We use 32 slices of thickness 6.4 Mpc. Our results are robust against  reasonable variation  of the slice thickness (not smaller than smoothing scale).

Fig.~\ref{fig:xH_slice} show the redshift evolution of a slice of the $x_{\rm HI}$ field plotted as functions of spatial coordinates $x,y$. These figures give essentially the same information as  Fig.~\ref{fig:xh_slice} but in a way that makes the visual interpretation of the CMT clearer. Note that the $z-$axis scales are not the same in different panels. We can see that at relatively early redshifts $z=18$ and 14 when the mean ionization values are close to one, at almost all threshold values there are numerous holes but only one connected region. At $z=9$ there are roughly the same number of holes and connected regions, and at $z=7$, there are mostly connected regions. In the following we will quantify how $\lambda_i$ and $\beta$ for each curve enclosing either a connected region or a hole, and their statistical behaviour, will inform us of the rate of progress of reionization. $x_{\rm HI}$ is highly skewed at high and low redshifts, as is evident from the panels showing $z=18, 14$ and 7.

We now define our notation for various quantities of interest.  We will choose threshold values, $\nu$, of the standard normal field $\tilde u$. $\nu$ will be in units of the standard deviation of $\tilde u$, which is one.
At each $\nu$ we denote the numbers of curves enclosing connected regions and  holes by $n_{\rm con}(\nu)$ and $n_{\rm hole}(\nu)$, respectively. The suffix `con' and `hole' refer to boundaries of connected regions and  holes, respectively. $n_{\rm con}(\nu)$ and  $n_{\rm hole}(\nu)$ are the Betti numbers for a random field~\cite{Chingangbam:2012,Park:2013}. Then, at each redshift $z$ we define,
\begin{equation}
  N_{\rm x}(z) \equiv \int_{\nu_{\rm low}}^{\nu_{\rm high}} {\rm d}\nu \,n_{\rm x}(\nu,z),
\label{eqn:Nx}
\end{equation}
where the suffix `x' stands for either `con' or `hole'. The lower and upper threshold cutoffs, $\nu_{\rm low}$ and $\nu_{\rm high}$, can be chosen depending on the threshold range of interest. For well behaved smooth random fields, $n_{\nu}$ goes to zero as  $\nu\pm\infty$. Therefore, the integral on the r.h.s of the above equation converges, and $N_{\rm x}$ is always finite even when the cutoff thresholds are taken to infinity. $N_{\rm x}(z)$ represents the ensemble of all curves within the chosen threshold range in the simulation box at a fixed redshift. Since for our practical purpose $\nu$ is sampled at a finite number of values, the integral is carried out as a Riemann sum.

We reserve the symbols $\lambda_i$, $r$ and $\beta$ to denote the eigenvalues, approximate radius and the ratio of the eigenvalues for a single curve. Let
\begin{eqnarray}
  {\overline\lambda}_{i,\rm x}(\nu) &\equiv& \frac{\sum_{j=1}^{n_{\rm x}(\nu)} \lambda_{i,\rm x}(j)}{n_{\rm x}(\nu)}, \\
  {\overline{r}}_{\rm x}(\nu) &\equiv& \frac{\sum_{j=1}^{n_{\rm x}(\nu)} r_{\rm x}(j)}{n_{\rm x}(\nu)}, \label{eqn:rx_nu}\\
  \quad {\overline\beta}_{\rm x}(\nu) &\equiv&  \frac{\sum_{j=1}^{n_{\rm x}(\nu)} \beta_{\rm x}(j)}{n_{\rm x}(\nu)},
  \label{eqn:betax_nu}
\end{eqnarray}
  denote their averages over all curves at a given $\nu$. Note that the usual scalar Minkowski Functionals are traditionally presented in this form as functions of the threshold. Then, we define 
\begin{eqnarray}
\lambda^{\rm ch}_{i,\rm x}(z) &\equiv& \frac{\int_{\nu_{\rm low}}^{\nu_{\rm high}} {\rm d}\nu \,n_{\rm x}(\nu,z) {\bar{\lambda}}_{i,\rm x}(\nu)}{N_{\rm x}(z)}, \\
 r^{\rm ch}_{\rm x}(z) &\equiv& \frac{\int_{\nu_{\rm low}}^{\nu_{\rm high}} {\rm d}\nu \,n_{\rm x}(\nu,z) {\bar{r}}_{\rm x}(\nu)}{N_{\rm x}(z)}, \label{eqn:rch}\\
\beta^{\rm ch}_{\rm x}(z) &\equiv& \frac{\int_{\nu_{\rm low}}^{\nu_{\rm high}} {\rm d}\nu \,n_{\rm x}(\nu,z) {\bar{\beta}}_{\rm x}(\nu)}{N_{\rm x}(z)}.
\label{eqn:betach}
\end{eqnarray}
These integrals are convergent for the same reason as explained for $N_{\rm x}(z)$.

{\em Physical interpretation of $ r^{\rm ch}_{\rm x}$ and $\beta^{\rm ch}_{\rm x} $}:  The physical information encoded in $ r^{\rm ch}_{\rm x}(z)$ and $\beta^{\rm ch}_{\rm x}(z) $ are essentially derived from $\lambda_i$. In the subsequent part of the paper we will present physical interpretation only for $r^{\rm ch}_{\rm x}(z)$ and $\beta^{\rm ch}_{\rm x}(z)$. 
$r^{\rm ch}_{\rm x}(z)$ condenses the size information for each type of structure from the entire field into one number, at each redshift. This number gives the largest possible average radius of the structure. It is well defined and independent of the observer measuring it. We interpret $r^{\rm ch}_{\rm x}(z)$ to be the {\em characteristic radius} for each type of structure, with the superscript `ch' referring to `characteristic'.  
In particular, $r^{\rm ch}_{\rm hole}$ for the $x_{\rm HI}$ field give the characteristic size of the ionized bubbles at each redshift. Likewise, we interpret $\beta^{\rm ch}_{\rm x}(z)$ to be the {\em characteristic shape anisotropy} of the corresponding structures at each redshift. 
Note that the results for the characteristic radius and shape will depend on the choice of the values of the threshold cutoffs. We will comment on this when we present our results.

\subsection{Results for $x_{\rm HI}$}
\label{sec:xH_results}
\begin{figure*}
\includegraphics[height=5.2in,width=5.9in]{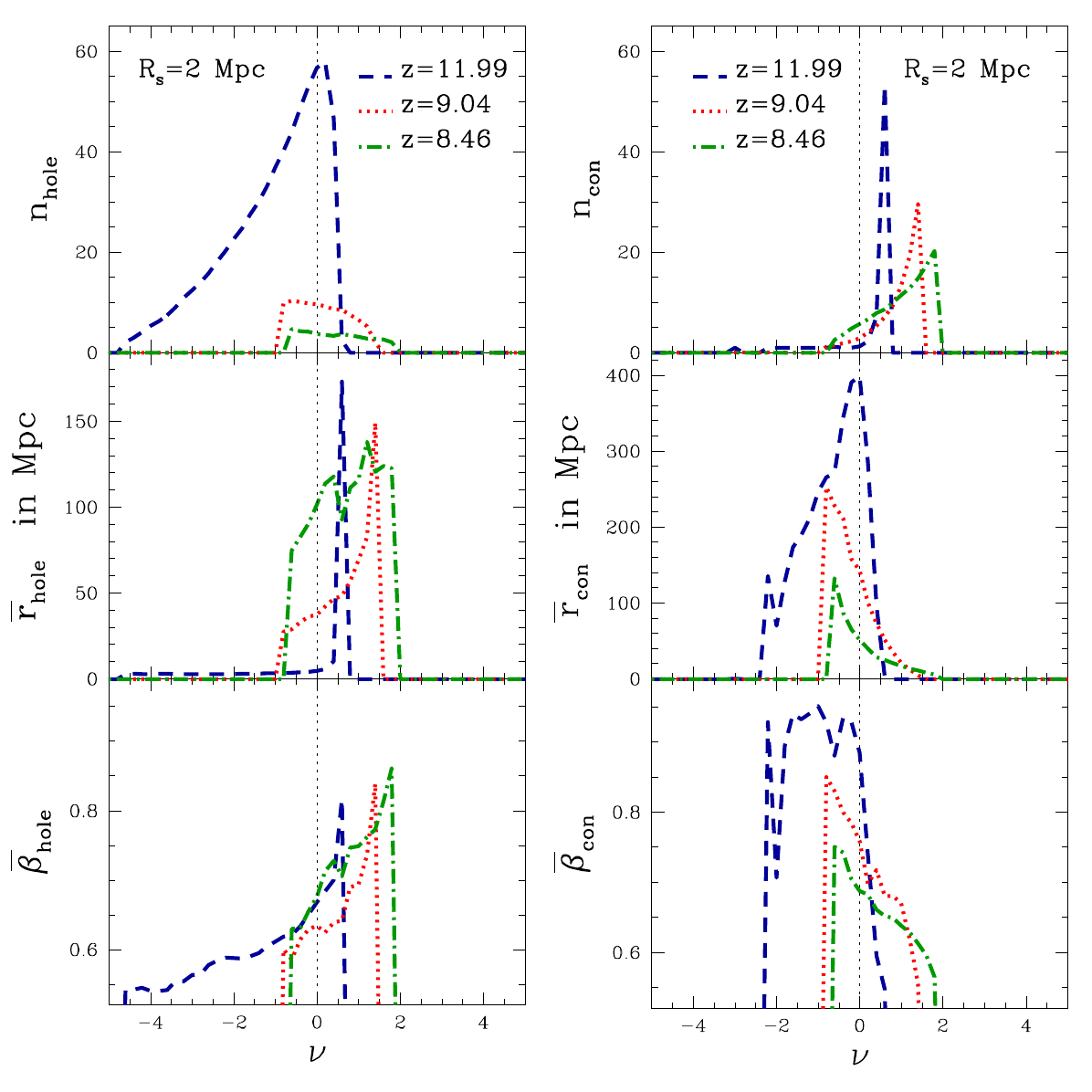}
\caption{Plots for $n_{\rm x}$ (top), ${\overline r}_{\rm x}$ (middle) and ${\overline\beta}_{\rm x}$ (bottom) versus $\nu$ (see Eqs.~\ref{eqn:rx_nu} and \ref{eqn:betax_nu}), for smoothing scale $R_s=2$ Mpc for different redshift values. Left column shows holes while right column shows connected regions.}
\label{fig:xH_vs_nu}
\end{figure*}

Let $y_{\rm HI}\equiv (x_{\rm HI}-\bar{x}_{\rm HI})/\sigma_{\rm HI}$, where $\sigma_{\rm HI}$ is the rms of  $x_{\rm HI}$. In Fig.~\ref{fig:xH_vs_nu} we show plots of $n_{\rm x}, {\overline r}_{\rm x}$ and ${\overline\beta}_{\rm x}$ versus $\nu$, (defined in Eqs.~\ref{eqn:rx_nu} and~\ref{eqn:betax_nu}) calculated using $y_{\rm HI}$.  The smoothing scale is $R_s=2$ Mpc.  The left column corresponds to holes, while the right corresponds to connected regions. All panels show a shift of the plots towards higher positive threshold range as the redshift decreases. This is due to decrease of $\overline x_{\rm HI}$. We see that at $z=11.99$ the number of holes (top, left panel) is large, spans a wide range of threshold values, and peaks at roughly the field mean value (correponding to $\nu=0$). The number of holes drops rapidly as the redshift decreases.  The sizes of holes (middle, left panel) are small for most values of $\nu$ at $z=11.99$. They are large at threshold values that correspond to field values close to $\overline x_{\rm HI}=1$. This simply means that, for level sets at such thresholds, holes  enclose large inter-connected regions which are highly non-convex. This correlates with the fact that the number of holes are few at such thresholds. As the redshift decreases, the size of holes show rapid increase. The shape of of holes (bottom, left panel) are more anisotropic at highly negative values and less so near threshold corresponding to $x_{\rm HI}=1$ for $z=11.99$. This implies that the shape of the random and highly non-convex level set boundary contours tend to be more isotropic, in comparison to that of the typically convex holes that are obtained for large negative threshold values. In general, ionized regions correspond to holes for negative threhold values. A precise identification of the ionized region, and its size and shape, will be threshold dependent. 

Connected regions have quite different redshift evolution in comparison to holes, as can be seen in the right column of Fig.~\ref{fig:xH_vs_nu}. They are more numerous at higher positive threshold values at all redshifts (top, right panel). They exhibit overall decrease of size with decreasing redshift (middle, right panel). We can also see that they are more anisotropic at higher thresholds (bottom, right panel). Note that we have chosen to show plots for $z=11.99,\ 9.04,$ and $8.46$ just so that the evolution trend for all quantities is easy to compare visually.  It is straightforward to extrapolate the trend to higher and lower redshift values. Complementary to holes, neutral regions correspond to connected regions for positive threhold values. A precise identification of the neutral region, and its size and shape, will again be threshold dependent. 

Although we do not focus on analyzing the non-Gaussian nature of the fields in this paper, we point out that the shapes of all the plots in  Fig.~\ref{fig:xH_vs_nu} reveal the highly non-Gaussian nature of the $x_{\rm HI}$ field. For isotropic Gaussian random fields, the expected shape of $n_{\rm x}$ as functions of the threshold have been worked out in~\cite{Chingangbam:2012,Park:2013}, while the shape of the matrix elements of $\mathcal W_1$ and $\beta$ have been calculated in~\cite{Vidhya:2016,Chingangbam:2017,Appleby:2017uvb}. These quantities will be very useful for investigating non-Gaussianity of the fields of the EoR.

\begin{figure*}
 \begin{center}
   \resizebox{2.5in}{2.5in}{\includegraphics{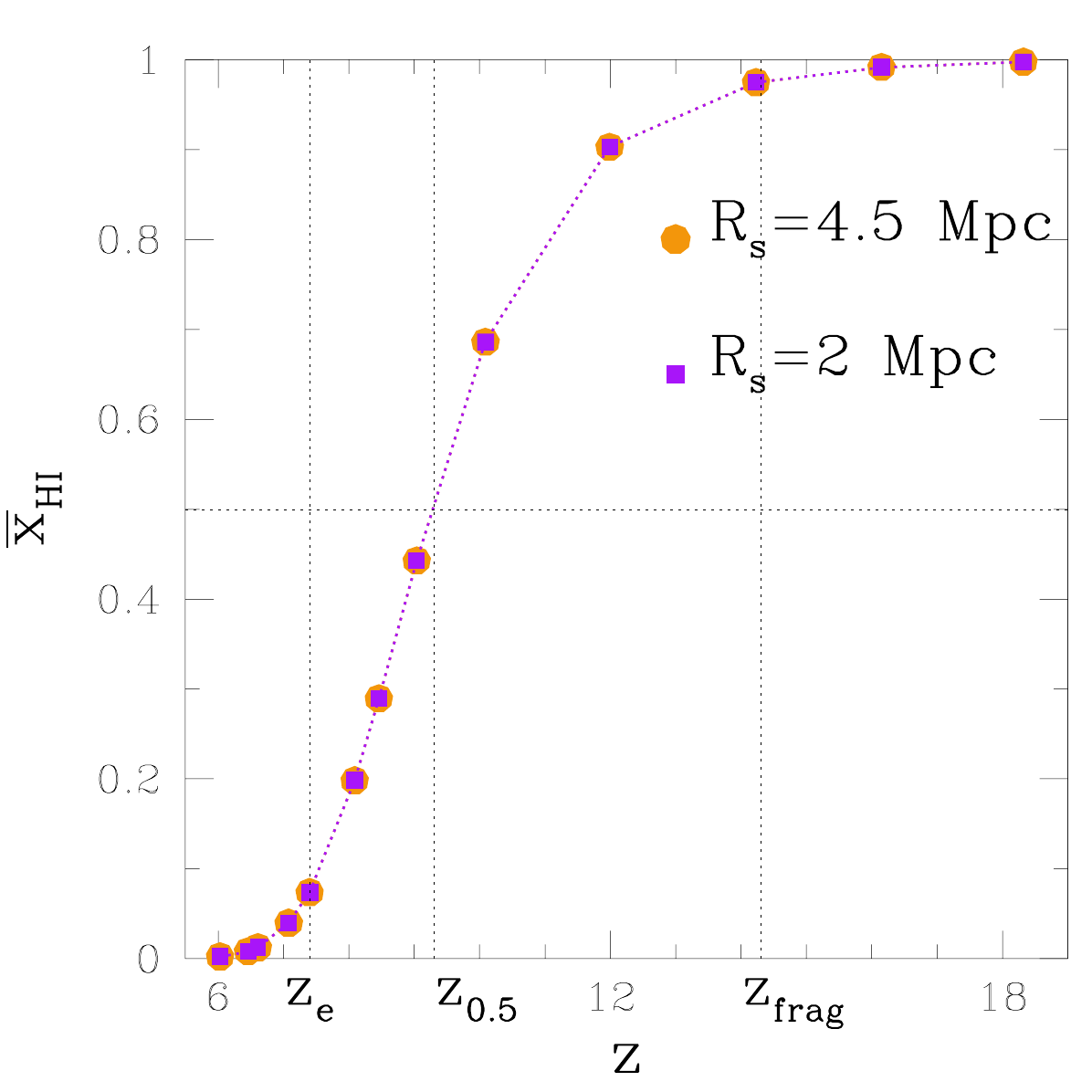}}
   \hskip 1.2cm
   \resizebox{2.5in}{2.5in}{\includegraphics{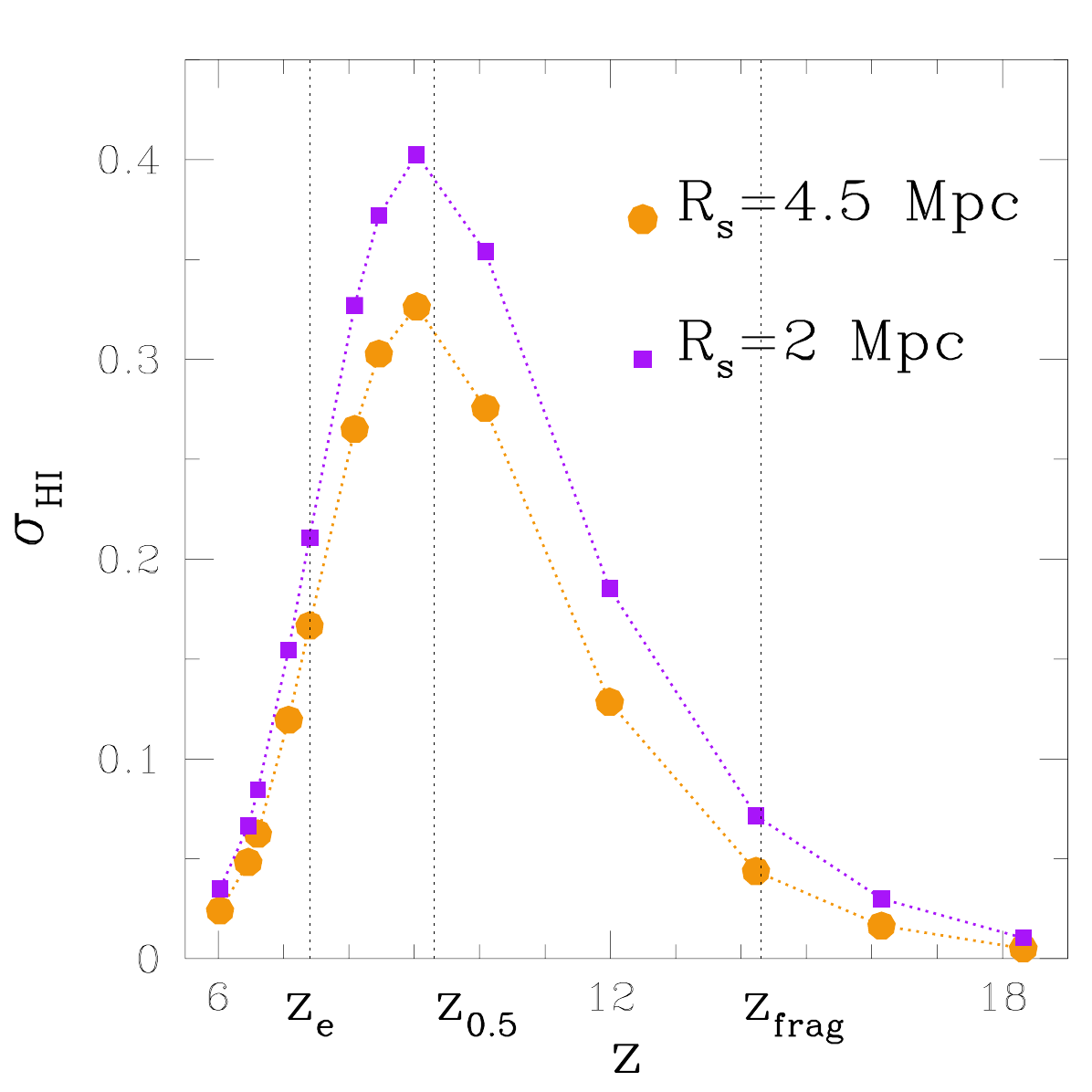}}\\  
   \resizebox{3.4in}{3.6in}{\includegraphics{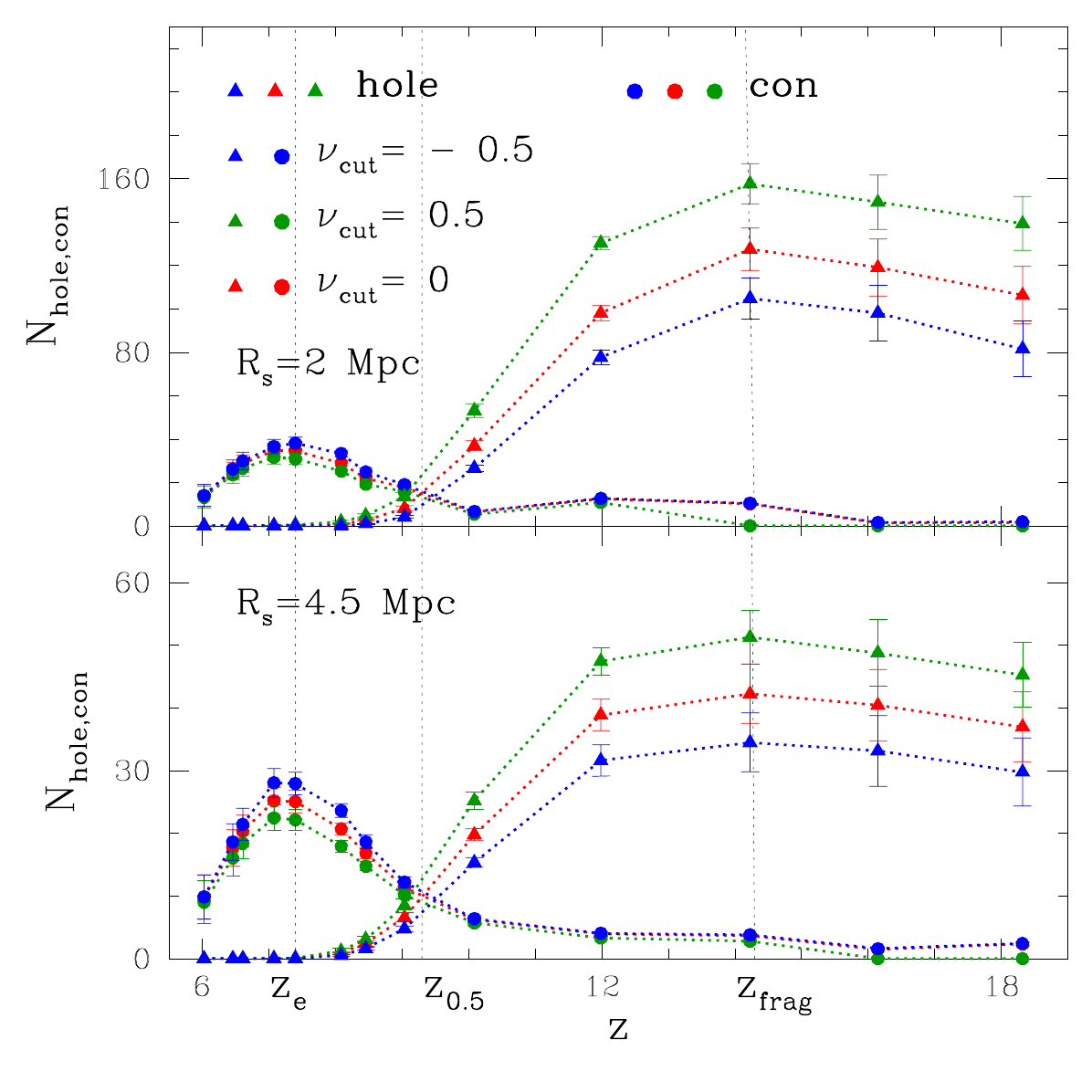}}
      \resizebox{2.4in}{2.8in}{\includegraphics{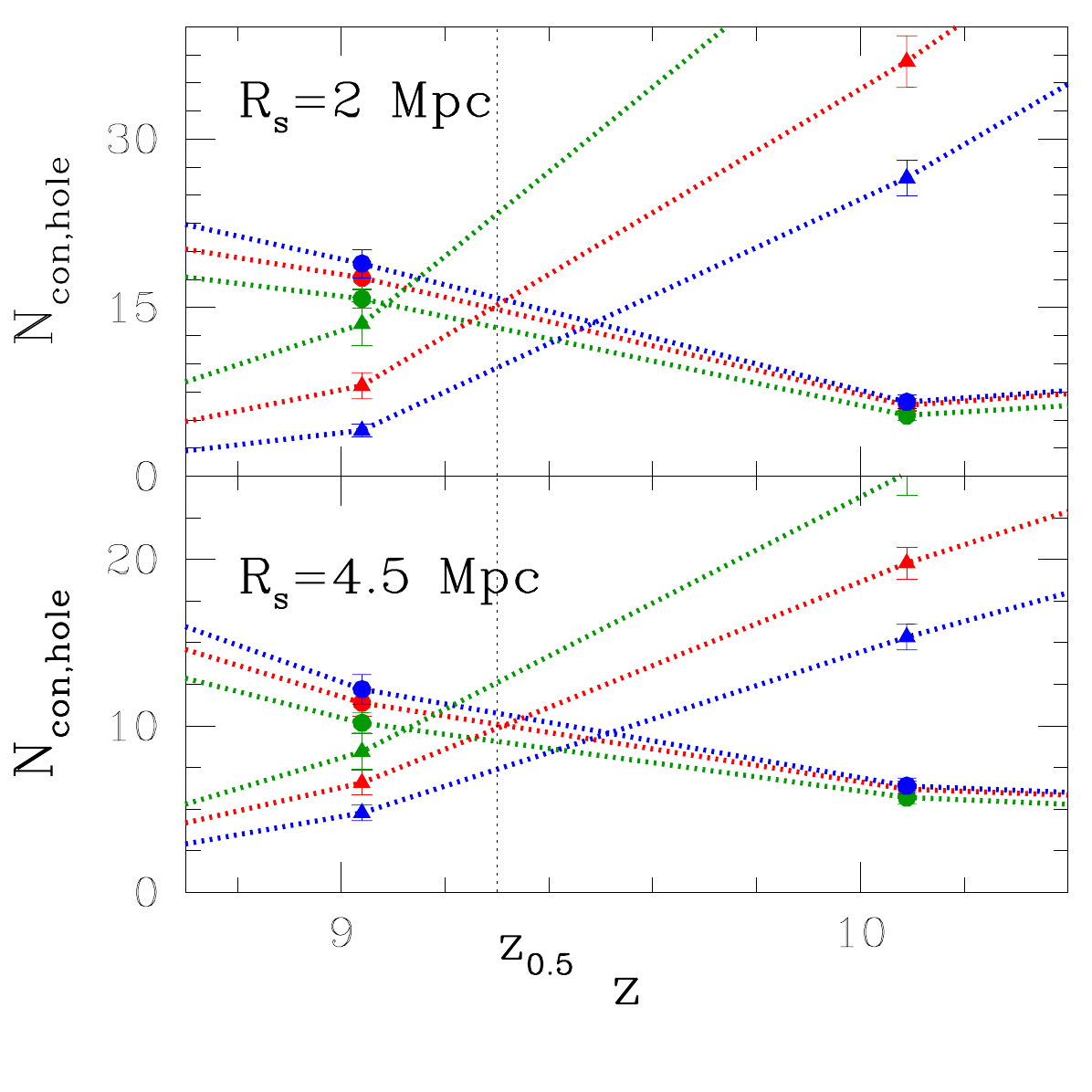}}
 \end{center}
 \caption{{\em Top left:} Redshift evolution of the mean of the ionization field, $\bar{x}_{\rm HI}$. {\em Top right:} Redshift evolution of the standard deviation of $x_{\rm HI}$, denoted by $\sigma_{\rm HI}$. {\em Bottom left:} $N_{\rm x}$ defined in Eq.~\ref{eqn:Nx} for $x_{\rm HI}$ for two smoothing scales $R_s=4.5$ and 2 Mpc, and three different choices of $\nu_{\rm cut}$. Dots represent connected regions while triangles represent holes. The three dotted vertical lines mark $z_{\rm frag} \sim 14.3$, $z_{0.5}\sim 9.3$ and $z_{\rm e}\sim 7.4$. Note that the $y-$axis ranges are different for the two values of $R_s$. {\em Bottom right:} Same as bottom left, but with the region of cross over between $N_{\rm con}$ and  $N_{\rm hole}$ zoomed in to highlight that $N_{\rm hole}$ and $N_{\rm con}$ cross over at $z=z_{0.5}$ for $\nu_{\rm cut}=0$. Error bars are the standard deviation from the 16 field slices.}
 \label{fig:xH_Nx}
\end{figure*}

\begin{figure*}
\includegraphics[height=5.2in,width=5.9in]{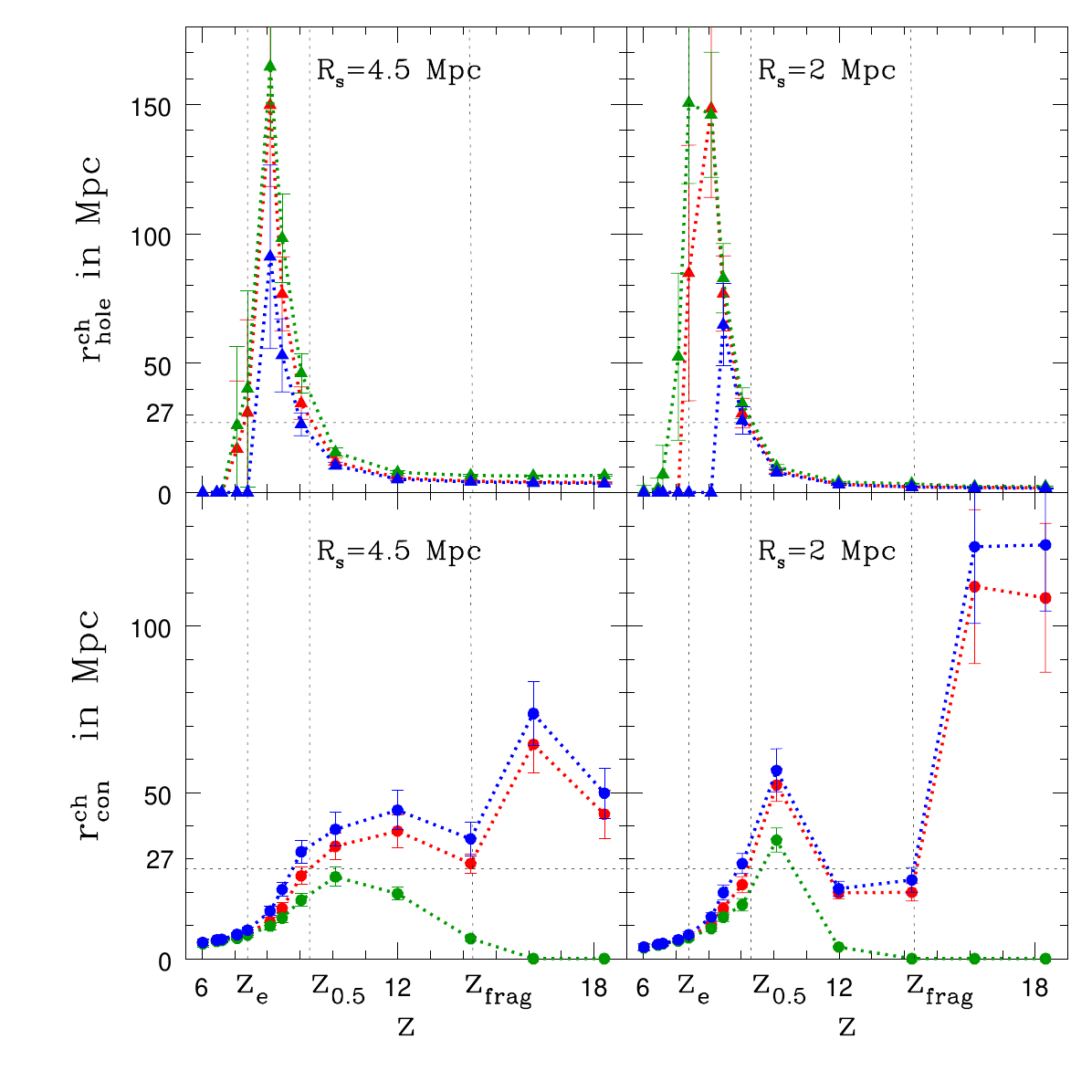}
 \caption{$r^{\rm ch}_{\rm hole}$ and $ r^{\rm ch}_{\rm con}$ (defined in Eq.~\ref{eqn:rch}),  for $x_{\rm HI}$ for two smoothing scales $R_s=4.5$ and 2 Mpc. Green, red and blue plots correspond to $\nu_{\rm}=0.5, 0$ and $-0.5$, respectively.  
   $z_{\rm frag}$, $z_{0.5}$ and $z_{\rm e}$ are marked. The horizontal dotted line marks the value of $ r^{\rm ch}_{\rm hole}$ at $z_{\rm e}$ for the case of  $\nu_{\rm cut}=0$. Error bars are the standard deviation from the 32 field slices.}
 \label{fig:xH_rc}
\end{figure*}

As mentioned above, the interpretation of the size and shape of ionized and neutral regions depends on the threshold of interest. In what follows, we choose to interpret them by computing the integrated quantities defined in Eqs. \ref{eqn:rch} and \ref{eqn:betach} so as to condense the information in the entire threshold range for each variable into a single number. 
For holes,  $\nu_{\rm low}$ is taken to be the minimum  value of $y_{\rm HI}$, and  $\nu_{\rm high}=\nu_{\rm cut}$, where  $\nu_{\rm cut}$ is chosen to be some appropriate value. For connected regions, $\nu_{\rm high}$ is taken to be the maximum  value of $y_{\rm HI}$, and  $\nu_{\rm low}=\nu_{\rm cut}$. We choose the value of $\nu_{\rm cut}$ to be around the mean value of $y_{\rm HI}$, which is zero. In order to show how the choice of $\nu_{\rm cut}$ affects the interpretation of number, shape and size of holes and connected regions we present calculations using $\nu_{\rm cut}=0,\pm 0.5$, at each redshift.  
These choices are made such that holes can be interpreted as ionized regions and connected regions as neutral regions. 

In the top panels of Fig.~\ref{fig:xH_Nx} we show the redshift evolutions of $\bar{x}_{\rm HI}$ (top left) and $\sigma_{\rm HI}$ (top right) for the smoothing scales $R_s=2$ and 4.5 Mpc, for the model of reionization that we are considering here. $z_{0.5}\simeq 9.3$ is marked by a vertical dotted line. The other two lines mark $z_{\rm frag}$ and  $z_{\rm e}$, which were defined earlier in section 2.1.  We can see that $\sigma_{\rm HI}$ is larger for smaller smoothing scale. It has a maximum at roughly $z_{0.5}$, and the peak location is not affected by smoothing scale.

$N_{\rm hole,con}$ are shown in the bottom left panel of Fig.~\ref{fig:xH_Nx}.  The bottom right panel shows a zoomed in version.   
Different colours correspond to different values of $\nu_{\rm cut}$: red for 0, blue for $-0.5$ and green for $+0.5$.  Triangles represent holes, while circles represent connected regions. 
$N_{\rm hole}$ exhibits two stages of evolution, namely, a slow increase at early redshifts which is then followed by a rapid decrease. The turnover takes place at around $z=z_{\rm frag}\simeq 14.3$.  $N_{\rm hole}$ shows a decrease as $\nu_{\rm cut}$ is decreased, which is as expected from its definition. In contrast, $N_{\rm con}$  first exhibits an increasing phase, followed by a decreasing phase. The turnover takes place at redshift $z=z_{\rm e}\simeq 7.3$. $N_{\rm con}$ shows an increase as $\nu_{\rm cut}$ is decreased, which is again as expected.  $N_{\rm hole}$ and $N_{\rm con}$ cross over at $z=z_{0.5}$, for $\nu_{\rm cut}=0$. As is clear from the plots, $z_{\rm frag}$ and $z_{\rm e}$ are independent of the smoothing scale and $\nu_{\rm cut}$. 
$z_{\rm frag}$ corresponds to $\bar x_{\rm HI}=0.97$ and $z_{\rm e}$ to $\bar x_{\rm HI}=0.04$. We note that the values of  $z_{\rm frag}$, $z_{0.5}$ and $z_{\rm e}$ that we have quoted are approximate and not based on accurate determination. 

Based on the evolution of $N_{\rm hole}$ and $N_{\rm con}$, we can divide the epoch of reionization into three time regimes, namely, $z \gtrsim z_{\rm frag}$, $ z_{\rm frag} \gtrsim z \gtrsim z_{\rm e}$, and $z \lesssim z_{\rm e}$.   
In the following we describe the behaviour of   $N_{\rm x}$,  $r^{\rm ch}_{\rm x}$ and $\beta^{\rm ch}_{\rm x}$ in these three time regimes.

\begin{figure*}
 \includegraphics[height=5.2in,width=5.9in]{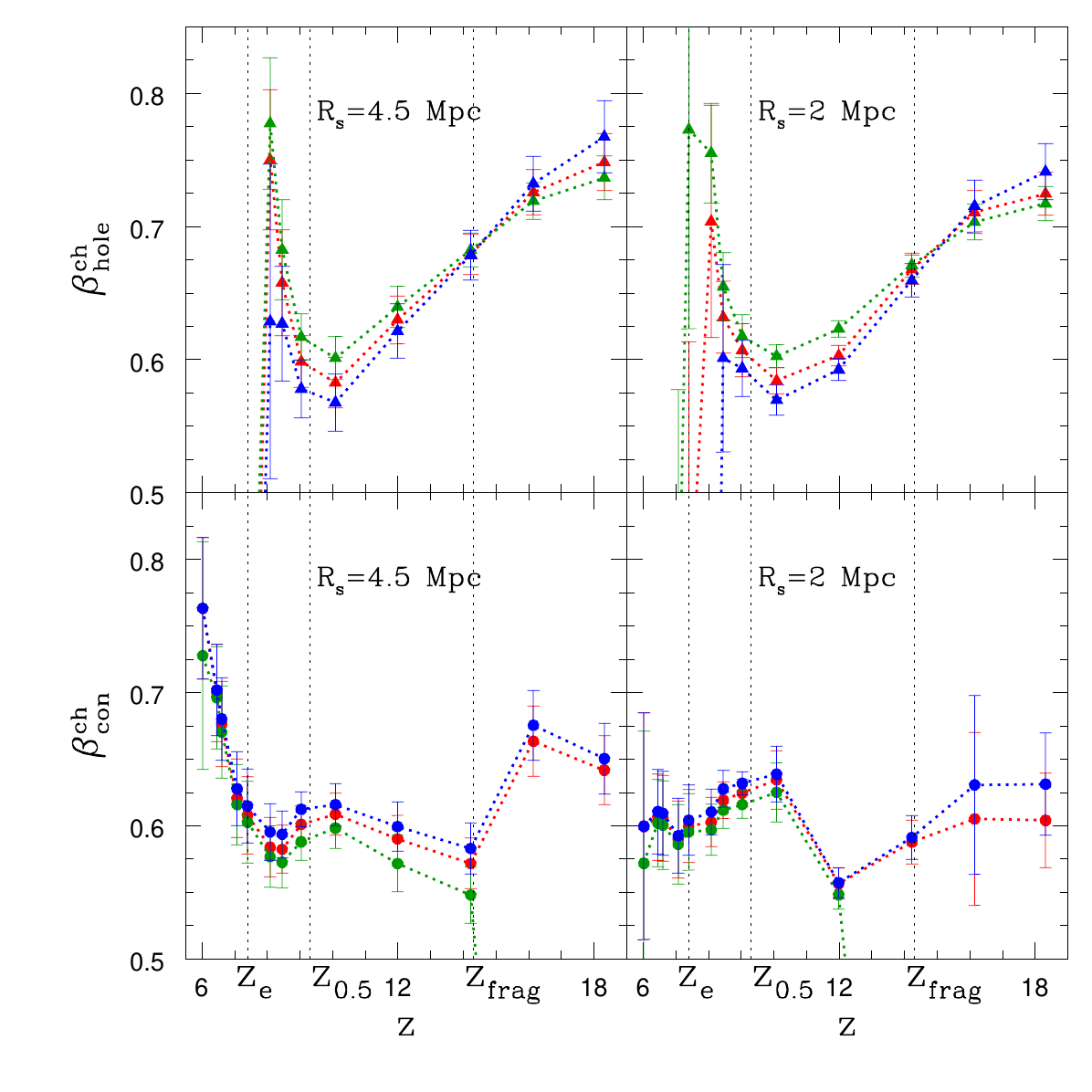}
 \caption{$\beta^{\rm ch}_{\rm hole}$ and $\beta^{\rm ch}_{\rm con}$ defined in Eq. \ref{eqn:rch} for $x_{\rm HI}$ for two smoothing scales $R_s=4.5$ and 2 Mpc. 
Green, red and blue plots again correspond to $\nu_{\rm}=0.5, 0$ and $-0.5$, respectively. Note that $\beta^{\rm ch}_{\rm x}=0$ means there are no structures. Error bars are the standard deviation from the 32 field slices.}
 \label{fig:xH_beta}
\end{figure*}

1. $z\gtrsim z_{\rm frag}$: During this epoch, ionized bubbles are being formed. 
The typical size of an ionizing source, when it emerges, is much smaller then the smoothing scales we choose.  Smoothing has the effect of smearing out the ionized field values. Hence the smoothed field is only partially ionized at every pixel. The value of $\bar{x}_{\rm HI}$ is close to one.  
The number of holes increases as more sources get formed,  and as a consequence $N_{\rm hole}$ increases with decreasing redshift till $z_{\rm frag}$.  The increase in $N_{\rm hole}$ indicates that the rate of formation of new ionizing sources happens faster than the rate of mergers of ionized regions. $N_{\rm con}(z)$ counts the isolated neutral regions. Since at this time the entire region is essentially one connected (neutral) regions with holes (ionized regions) puncturing it, the information in $N_{\rm con}(z)$ is not statistically significant. 

$r^{\rm ch}_{\rm hole,con}$ are shown in Fig.~\ref{fig:xH_rc}. Redshifts where $r^{\rm ch}_{\rm hole,con}=0$ means there are no structures. Different colors correspond to different values of $\nu_{\rm cut}$, as in Fig.~\ref{fig:xH_Nx}. $r^{\rm ch}_{\rm hole}$ is roughly constant during this epoch and is determined by the smoothing scale. During this time $r^{\rm ch}_{\rm hole}$ gives a good estimate of the size of ionized bubbles because most ionized regions have convex boundaries. In Fig.~\ref{fig:xH_slice} we notice that for early and late redshifts, level sets near the mean value of the field tend to have larger boundaries compared to other threshold values. As a consequence, 
larger $\nu_{\rm cut}$ results in larger $r^{\rm ch}_{\rm hole}$.

As mentioned earlier there is essentially just one connected region during this regime. Depending on the value of $\nu_{\rm cut}$ we can obtain more than one connected region due to the uneven nature of the field near its mean value. $r^{\rm ch}_{\rm con}=0$ for large redshifts for $\nu_{\rm cut}=0.5$ because the maximum value of the field $y_{\rm HI}$ is lower than this value. At $z_{\rm frag}$, as the neutral region begins fragmenting $r^{\rm ch}_{\rm con}$ becomes useful to probe.

$\beta^{\rm ch}_{\rm hole,con}$ are shown in Fig.~\ref{fig:xH_beta}.  At $z\simeq 18$ we find that $\beta^{\rm ch}_{\rm hole}$ is less than one, and larger for smaller $\nu_{\rm cut}$. This is because the ionized regions tend to be more isotropic close to the ionizing centres. Moreover, smaller smoothing scale gives smaller values of $\beta^{\rm ch}_{\rm hole}$. Therefore, even at such early redshifts ionized regions are not isotropic. They become more anisotropic as the redshift decreases. The increase of anisotropy implies that the rate of merger of ionized regions is nonzero and increasing (as depicted in Fig. \ref{fig:bubblemerger}), even though in this time regime the rate of formation of new ionizing sources is greater than the merger rate. 
We find that the plots of $\beta^{\rm ch}_{\rm hole}$ for the three values of $\nu_{\rm cut}$ cross just before $z_{\rm frag}$ when the merger rate overtakes the source formation rate.  
$\beta^{\rm ch}_{\rm con}$ also exhibits anisotropy during this regime. The average anisotropy is more than that of holes (lower value of $\beta$).   

2. $ z_{\rm frag} \gtrsim z \gtrsim z_{\rm e}$: $N_{\rm hole}$ drops sharply after $z_{\rm frag}$,  as seen in  Fig.~\ref{fig:xH_Nx} due to rate of mergers becoming larger than the rate of formation of sources. Fragmentation of the connected region results in increase of $N_{\rm con}$. At roughly $z_{0.5}$,  $N_{\rm con}\sim N_{\rm hole}$.
 After  $z_{0.5}$, $N_{\rm con}$ continues increasing as the fragmentation process carries on, while  $N_{\rm hole}$ continues to decrease. 

From $z_{\rm frag}$ onwards $r^{\rm ch}_{\rm hole}$ grows, while $r^{\rm ch}_{\rm con}$ decreases, as  seen in Fig. \ref{fig:xH_rc}. They eventually cross over, which means the average sizes of ionized and neutral regions are the same. Both $r^{\rm ch}_{\rm hole}$  and  $r^{\rm ch}_{\rm con}$ depend on $\nu_{\rm cut}$, as expected. For the case  of $\nu_{\rm cut}=0$ we find that their cross over happens at $z=z_{0.5}$. $r^{\rm ch}_{\rm hole}$  provides a fair estimate of the ionized bubble size until the bubble boundaries become highly non-convex due to multiple mergers. As mentioned in section 3, the area enclosed within the ionized regions is actually smaller than the area of the circle whose radius is given by  $r^{\rm ch}_{\rm hole}$. We find that $r^{\rm ch}_{\rm hole}$ grows to a value of 27 Mpc at $z_{0.5}$ for $\nu_{\rm cut}=0$ (horizontal dotted line). This number is independent of the smoothing scale, as can be expected since the smoothing scale is much smaller than this derived characteristic size. 

  We see in  Fig. \ref{fig:xH_rc} that $r^{\rm ch}_{\rm hole}$ continues to grow well beyond $z=z_{0.5}$. By this time the number of holes is very few,  and consists of small ionized regions enclosed within neutral regions and others that are very large and highly non-convex. The large non-convex ones result in the growth of $r^{\rm ch}_{\rm hole}$ that we find. Due to the high non-convexity of the holes it is possible that some holes have  boundary perimeters that are larger than the perimeter of the simulation box. 
As mentioned earlier, the high non-convexity of the holes during this time also implies that though the perimeter is obtained from exact calculation,  the interpretation of $r^{\rm ch}_{\rm hole}$ as the radius of an idealized circle over-estimates the actual size of the ionized regions to a large extent. Therefore, it is not a useful interpretation during this time regime. 
Meanwhile, $r^{\rm ch}_{\rm con}$ continues to drop as neutral regions fragment further.

We find that $\beta^{\rm ch}_{\rm hole}$ continues to decrease further after  $z_{\rm frag}$, and then exhibits a turn around at around $z\simeq 10$, just before $z_{0.5}$. The turn around redshift is independent of smoothing scale. 
This indicates that most bubbles have merged and the few holes contributing to the value $\beta$ are the highly non-convex ones close to the field mean value which tend to be more isotropic. In contrast, $\beta^{\rm ch}_{\rm con}$ exhibits increasing anisotropy after $z\simeq 10$, has a minimum a around $z\simeq 8.5$ and then tends to become isotropic after, for $Rs=4.5$ Mpc. The behaviour prior to $z\simeq 10$ is dominated by statistical fluctuation and is statistically insignificant. For  $Rs=2$ Mpc we do not see any rise in isotropy, which can be explained by the fact that larger Gaussian smoothing tends to isotropize small scale regions.

3. $z\lesssim z_{\rm e}$:   As the ionization process takes over most of the region, $N_{\rm con}$ will turn over and decrease, and we find that this happens at around $z_{\rm e}\simeq 7.4$. By this time the EoR is getting complete and both $N_{\rm con}$ and $N_{\rm hole}$ drop towards zero. $r^{\rm ch}_{\rm hole}$ peaks at before $z_{\rm e}$, depending on $\nu_{\rm cut}$. Thereafter it drops sharply as only a few holes are left within neutral regions. $r^{\rm ch}_{\rm con}$ continues to fall towards zero as the entire region becomes fully ionized. 

As was the case for connected regions in regime 1, in this regime all physical quantities associated with holes regions during this epoch are not statistically significant since the number of hole regions enclosed within neutral regions is practically zero. 

\subsection{Results for $\delta T_{\rm b}$}
\begin{figure*}
 \begin{center}
   \resizebox{6.2in}{5.3in}{\includegraphics{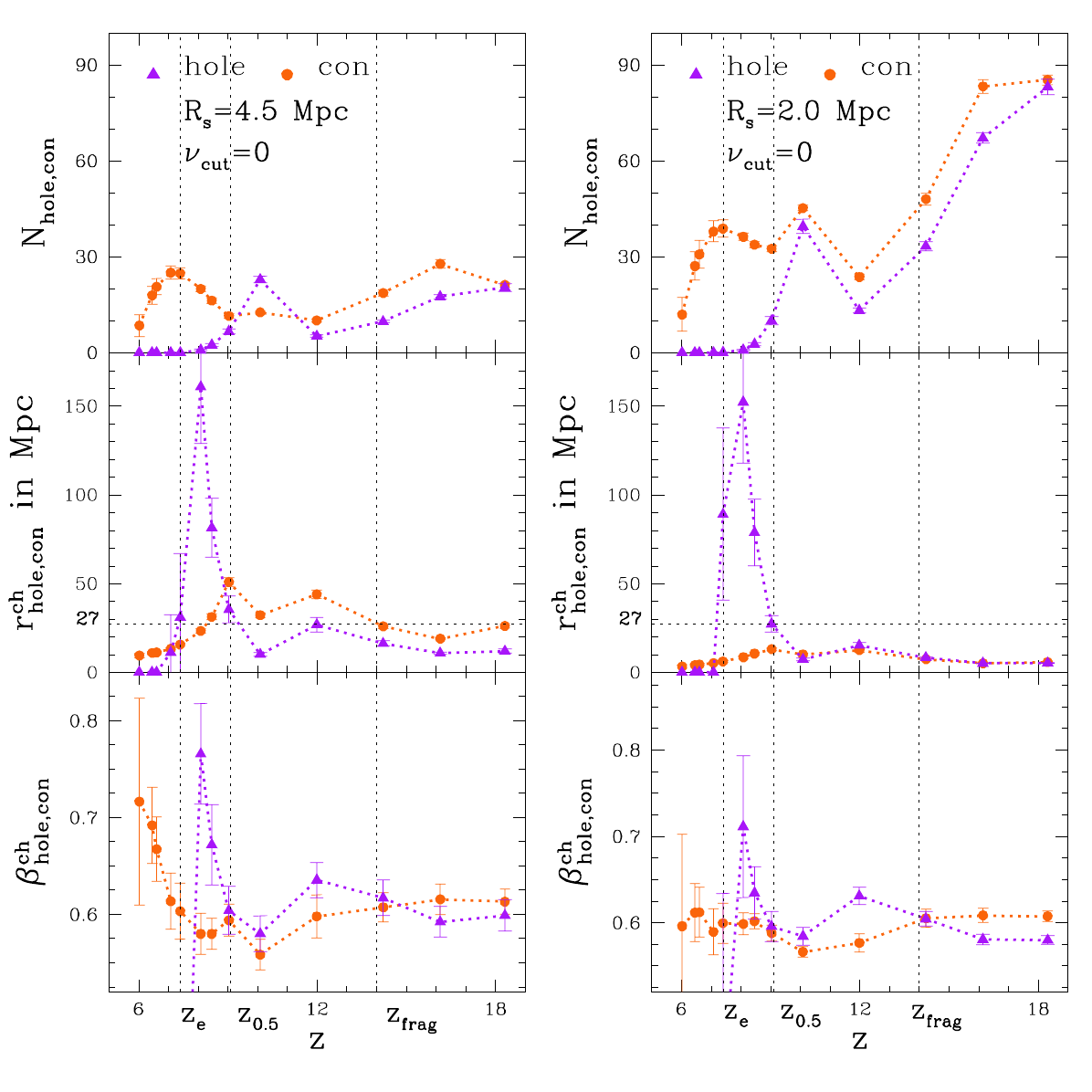}}
   \end{center}
 \caption{$N_{\rm x}$,  $r^{\rm ch}_{\rm x}$ and $\beta^{\rm ch}_{\rm x}$ for $\delta T_{\rm b}$, for the same smoothing scales as for $x_{\rm HI}$, for $\nu_{\rm cut}=0$. Purple triangles denote holes while orange dots denote connected regions. The three dotted vertical lines again mark $z_{\rm frag} \sim 14.3$, $z_{0.5}\sim 9.3$ and $z_{\rm e}\sim 7.4$ for comparison with Figs.~\ref{fig:xH_Nx}, \ref{fig:xH_rc} and \ref{fig:xH_beta}. Error bars are the standard deviation from the 16 field slices.}
 \label{fig:Tb_betti_r_beta}          
\end{figure*}

In Fig.~\ref{fig:Tb_betti_r_beta} we show the redshift evolutions of $N_{\rm x}$,  $r^{\rm ch}_{\rm x}$ and $\beta^{\rm ch}_{\rm x}$ for $\delta T_{\rm b}$ for two smoothing scales $R_s=4.5$ (left panels) and 2 Mpc (right panels). Orange dots show connected regions and purple triangles show holes. 
We can see that the redshift evolutions for all quantities trace that of $x_{\rm HI}$  at later redshifts from roughly $z\sim 10$ onwards. The physical reason for this behaviour is that at these relatively low redshifts the factor that depends on $T_{\gamma}/T_s$ in the expression for $\delta T_b$ given in Eq.~\ref{eqn:dTb} becomes negligible due to $T_s$ becoming much larger than  $T_{\gamma}$.

For redshifts higher than $z>10$ the behaviour of  $\delta T_b$ will be primarily controlled by that of $T_s$. We postpone analysis of $T_s$ and $\delta_{\rm NL}$ to future work since our main interest here is to understand the ionization field.   

\subsection{Probability distributions of $r$ and $\beta$ for $x_{\rm HI}$ and $\delta T_b$}

We now estimate the probability distribution functions (PDF)  of $r$ and $\beta$. The correct way to analyze the PDF would be to calculate it for all structures at each threshold. However, our simulation box is not big enough to provide sufficiently large $n_{\rm x}(\nu)$ to obtain good statistical result. Hence, we use the ensemble of {\em all curves} at {\em all threshold values} sampled. As a result the sizes and shapes of the curves are not all independent, but those corresponding to one peak or trough of the field are correlated. Nevertheless, using the full ensemble of curves will reflect the true PDF. 

\begin{figure*}
 \begin{center}
   \resizebox{3.in}{3.in}{\includegraphics{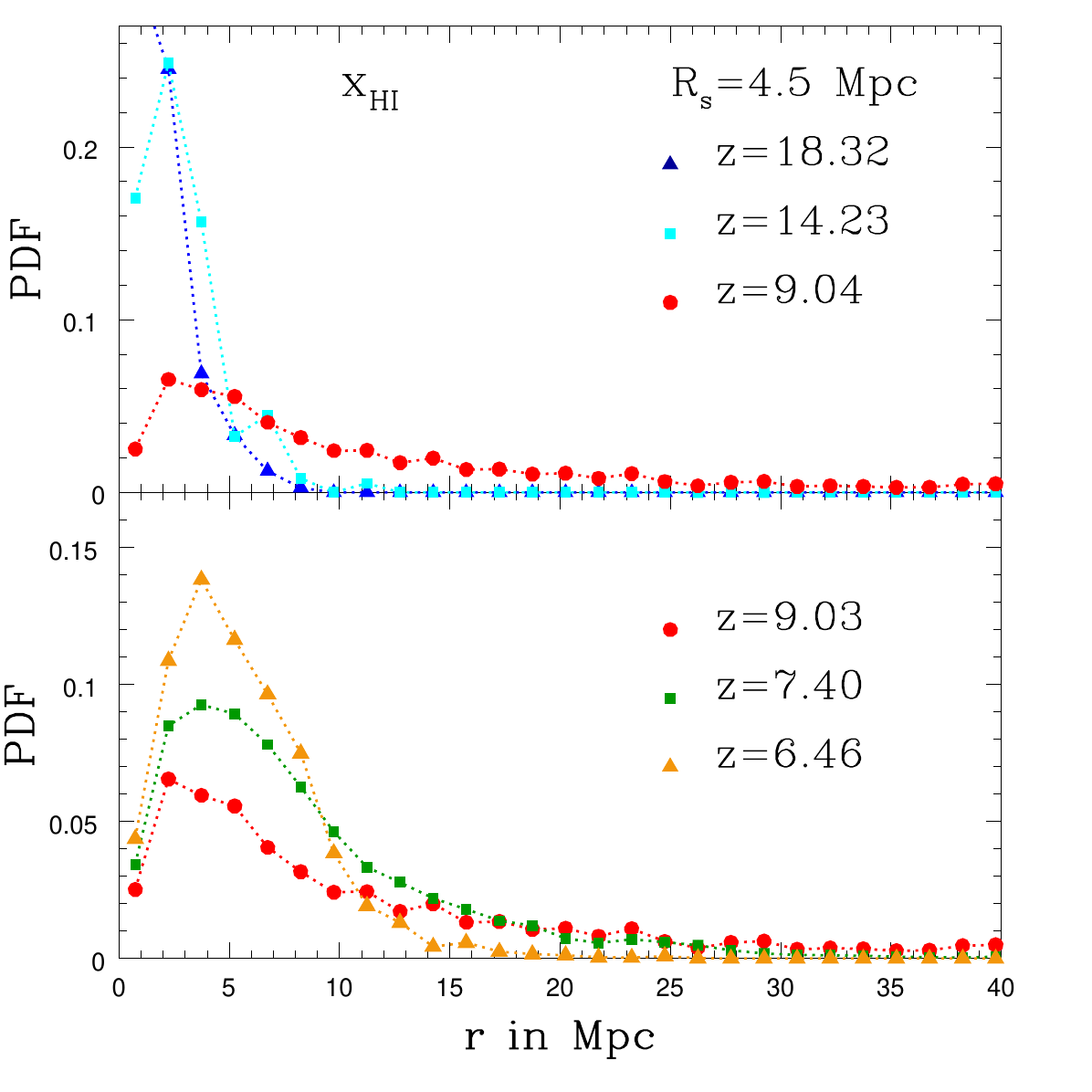}}
   \resizebox{3.in}{3.in}{\includegraphics{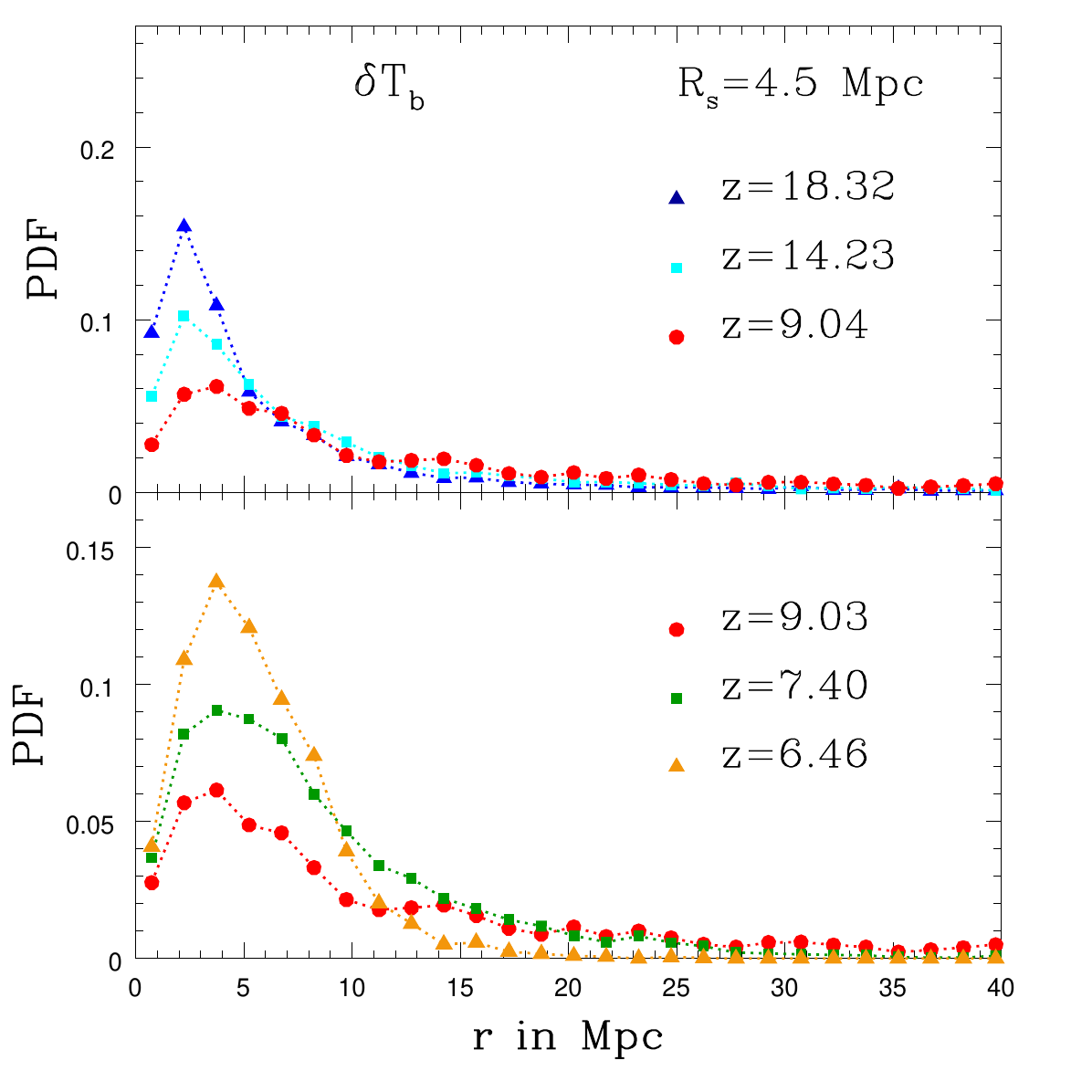}}\\
   \end{center}
 \caption{PDFs of the size parameter $r$ obtained from the ensemble of all curves, including both holes and connected regions, from all sampled threshold values, at different redshift for $x_{\rm HI}$ (left) and $\delta T_b$ (right). }
\label{fig:pdf_lambda}
\end{figure*}

In Fig.~(\ref{fig:pdf_lambda}) we show the PDFs of $r$ for $x_{\rm HI}$ (left panel) and $\delta T_b$ (right panel) for $R_s=4.5$ Mpc. We find that the PDFs are highly skewed, as can be expected. We have chosen to show $x$ range upto 40 Mpc so as to highlight the redshift evolution for lower values of $r$ where the probabilities are high. The tails extend well beyond 40 Mpc, particularly at intermediate redshifts around $z_{0.5}$. For $x_{\rm HI}$, we see that the PDFs in the upper panel become flatter as the redshift decreases from high values to the intermediate value of 9.04, indicating that the mean value is becoming larger. The structures are predominantly holes during this time and the increase of the mean reflects the growing size of the ionized regions. In the lower panel, we see that the probabilities again increase towards smaller values of $r$ as the redshift drops below 9.04. This is due to the structures during this time being predominantly fragmented connected regions which are becoming smaller as ionization progresses. In the right panels showing $\delta T_b$ we can see that the plots in the lower panel trace the corresponding plots for $x_{\rm HI}$, while those in the upper panels differ. This is again due to $\delta T_b$ tracing $x_{\rm HI}$ during the later redshifts.

\begin{figure*}
\begin{center}
      \resizebox{3.in}{3.in}{\includegraphics{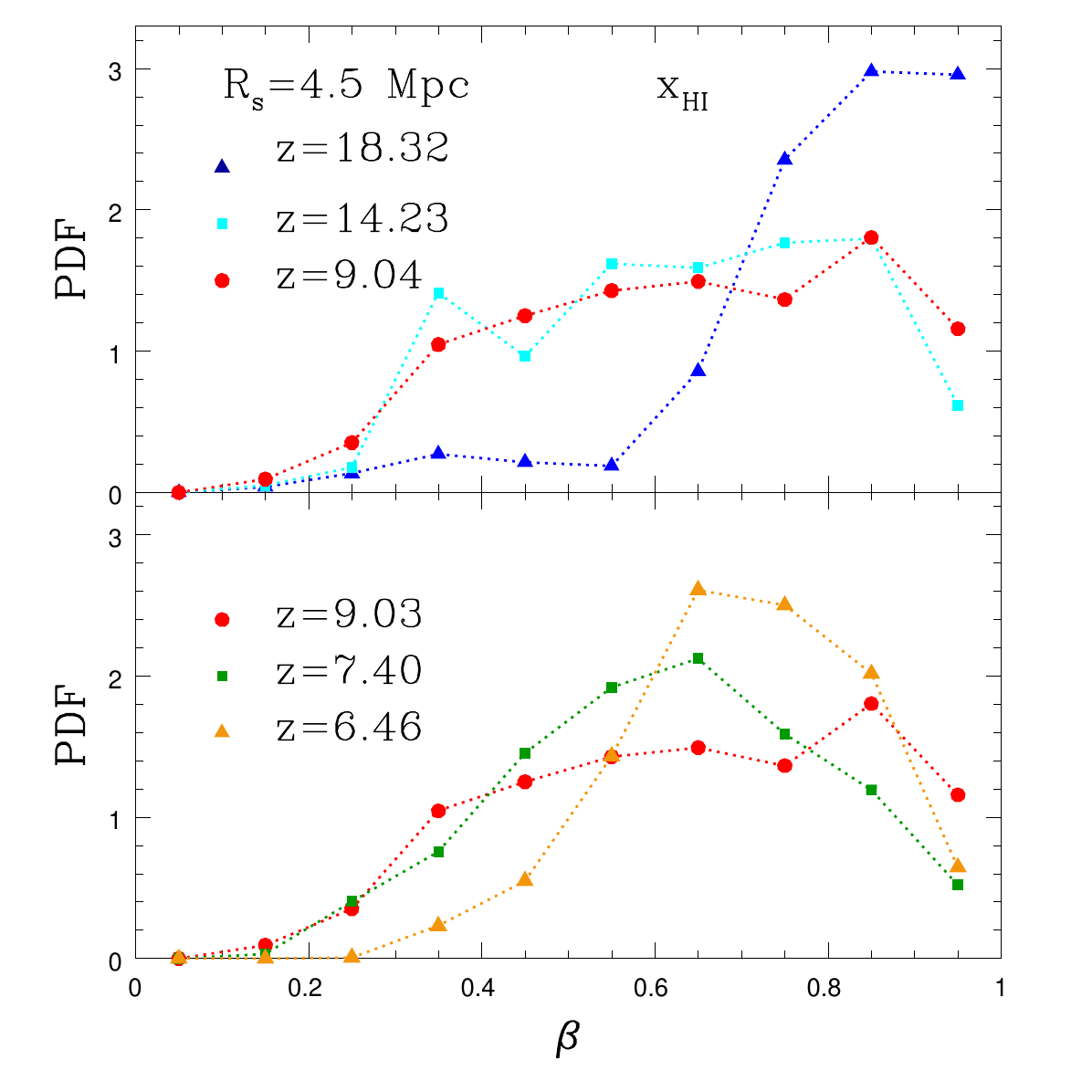}}
   \resizebox{3.in}{3.in}{\includegraphics{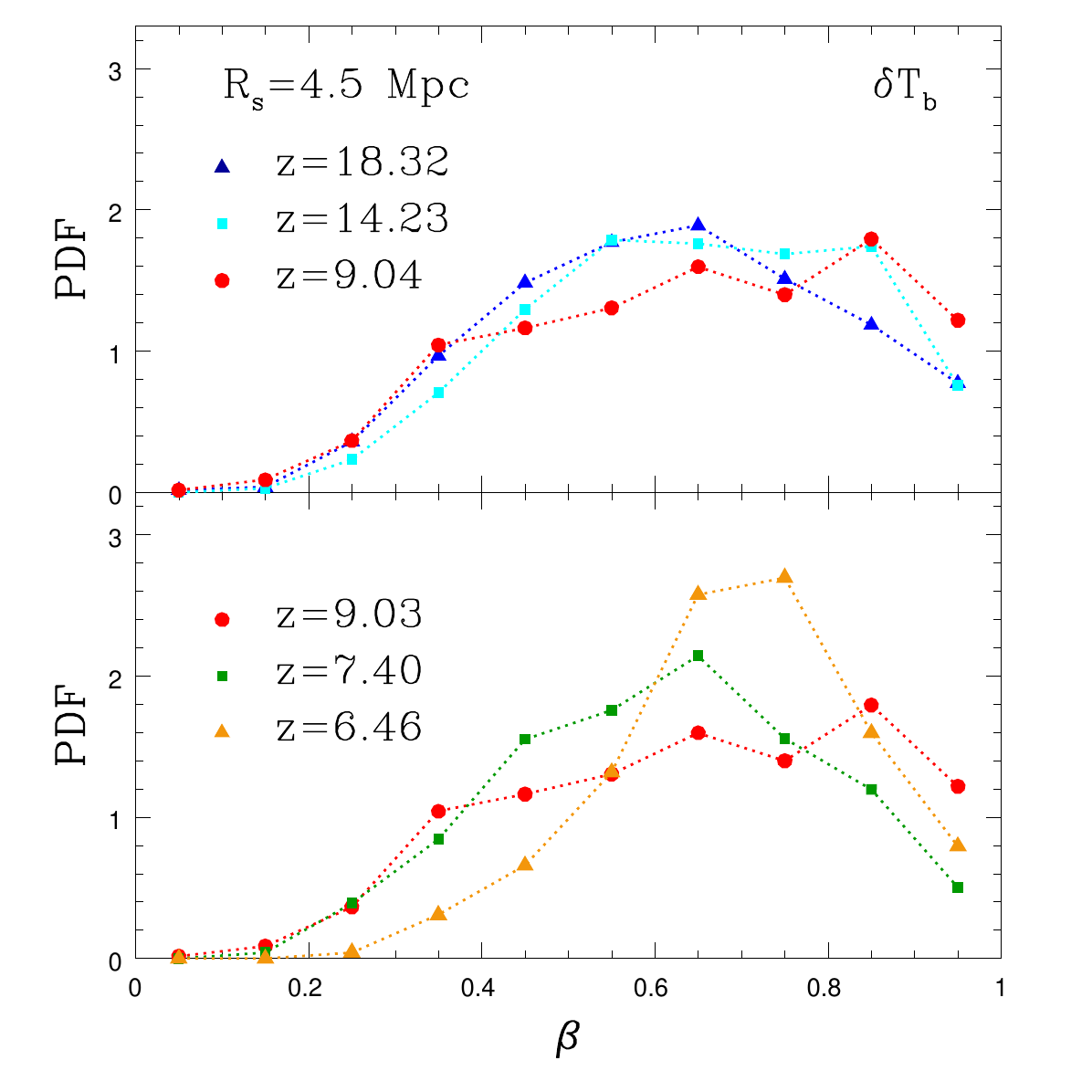}}
 \end{center}
 \caption{Redshift evolution of the PDF of the shape parameter $\beta$ obtained from the ensemble of all curves of our threshold sampling. The shift of the peak to lower $\beta$ values implies that as the redshift decreases to roughly $z\simeq z_{0.5}$ the shape of ionized bubbles become more anisotropic. This is due to merger of bubbles becoming numerous. Then, as $z$ decreases further below $z_{0.5}$, most mergers are over, and the PDF reflects the shape of connected (neutral) hydrogen regions, and their overall shape tends is less anisotropic.}
 \label{fig:pdf_beta}
\end{figure*}

In Fig.~(\ref{fig:pdf_beta}) we show the PDFs of $\beta$ for $x_{\rm HI}$ (left) and $\delta T_b$ (right), for the same redshifts and smoothing scale as in  Fig.~(\ref{fig:pdf_lambda}). All the  PDFs tend to zero as $\beta\rightarrow 1$, implying that exactly isotropic structures are improbable. 
In the top left panel for $x_{\rm HI}$ we can see that the peaks of the PDFs shift towards lower values of $\beta$ as the redshift decreases. This reflects the fact that the mean shape of ionized bubbles becomes more anisotropic, as explained in section 4.1. Then in the bottom left panel we see that the peaks of the PDFs again shift back to higher values of $\beta$ as the redshift decreases, and the turn over takes place around $z\sim z_{\rm 0.5}$. The PDFs of $\delta T_b$ on the right again trace those of $x_{\rm HI}$  for lower redshifts and differ for higher redshifts. 

\section{Conclusion and discussion}
\label{sec:conclusion}

We have proposed in this paper a new method based on the contour Minkowski Tensor to probe the shape and size statistics of ionized bubbles. These statistics reveal important time scales associated with the EoR. 
Our first result is that ionized bubbles are not isotropic in shape, as can be expected from visual inspection of simulations of the ionization fraction field, and our method gives a precise quantification of the anisotropy.
Our second result is the quantification of the mean size of ionized bubbles and their redshift evolution. For the model of reionization considered in this paper, we find that the characteristic radius of ionized bubbles grow to the value 27 Mpc at $z_{0.5}$, independent of smoothing scales. We show that the variation of the size and shape parameter as a function of redshift reveal the time epochs when mergers of ionized bubbles become dominant over the creation of new ionizing sources, when maximum mergers happen, and  when mergers end due to the entire region becoming ionized.

As shown in section \ref{sec:xH_results}, the value of the bubble size varies with the choice of the limits of integration over the standard normal field threshold. However, we emphasize that upon fixing the integration limits the result is unambiguous and can be  compared meaningfully across different simulations (and observed data in the future) so as to understand the impact of variation of physical parameters of different models of the EoR. For the model that we have used in this work, our result is exact, upto the numerical inaccuracies of the Riemann sum that approximate the integrations. Further we have presented our result in the form of a radius of an idealized circle, in keeping with the usual discussions found in the literature. However, {\em given the highly complicated shapes of the ionized regions it makes better sense to use the perimeter of the regions directly as a physical observable rather than an idealized radius}. It is also worth mentioning that a precise calculation of the mean area enclosed by the ionized bubbles boundaries is feasible by using the area fraction, which is the first Minkowski functional, together with the Betti numbers. We will carry out this calculation in the near future.  We also plan to do a detailed comparison of the bubble size statistics derived using our method with other methods in the literature. The extension of our work to three dimensional MTs is ongoing.

The calculations shown in this paper represent the first step showing  the usefulness of the CMT in probing the EoR. We have concentrated on developing the technique and applied it mainly to the ionization field. Hence we have not done justice to a full analysis of the physics that operated during the EoR. We have left out analysis of the matter field and spin temperature which play important roles at early redshifts. We have also ignored the effect of peculiar velocity. 
We have demonstrated our results for only one history of reionization or model. The details of the redshift evolution - the precise shape of $r^{\rm ch}_{\rm x}$  and $\beta^{\rm ch}_{\rm x}$ versus $z$, and the location of the transition time scales are expected to vary depending on the details of the reionization process. Thus, we expect that our method will be useful to test different models of reionization. We plan to carry out these analyses in our follow up work and a  comparative study of different models is ongoing. We reiterate that our analysis is based on idealized simulations, and the reliability and usefulness  of our method in realistic observational scenarios where foreground contamination and instrument noise dominate the signal, needs to be investigated further. Moreover, the error bars that we have shown are obtained using two-dimensional slices from the three-dimensional simulations. The correlation that is inherent due to the slicing implies that our error bars are underestimated. 
 
\acknowledgments{
The computation required for this work was carried out on the Hydra
cluster at the Indian Institute of Astrophysics. The code for computing the CMT was developed on the Center for Advanced Computation, Korea Institute for Advanced Study. 
We acknowledge use of the \texttt{21cmFAST} code~\cite{Mesinger:2010ne}.
P.~C would like to thank K.~P.~Yogendran for useful discussions.}


\begin{thebibliography}{99}


\bibitem{Madau:1997} P.~Madau, A.~Meiksin and M.~J.~Rees, 
  Astrophys.\ J.\  {\bf 475}, 429 (1997)
  
\bibitem{Furlanetto:2006jb} 
  S.~Furlanetto, S.~P.~Oh and F.~Briggs,
  Phys.\ Rept.\  {\bf 433}, 181 (2006)

\bibitem{Cen:2005ax} 
  R.~Cen,
  [astro-ph/0507014].

\bibitem{Wyithe:2004qd} 
  J.~S.~B.~Wyithe and A.~Loeb,
  Nature {\bf 432}, 194 (2004)
  doi:10.1038/nature03033
  [astro-ph/0409412].
  
 \bibitem{Furlanetto:2004nh} 
  S.~Furlanetto, M.~Zaldarriaga and L.~Hernquist,
  Astrophys.\ J.\  {\bf 613}, 1 (2004)

\bibitem{Furlanetto:2005ax} 
  S.~R.~Furlanetto, M.~McQuinn and L.~Hernquist,
  Mon.\ Not.\ Roy.\ Astron.\ Soc.\  {\bf 365}, 115 (2006)
  doi:10.1111/j.1365-2966.2005.09687.x
  [astro-ph/0507524].
  
\bibitem{McQuinn:2006et} 
  M.~McQuinn, A.~Lidz, O.~Zahn, S.~Dutta, L.~Hernquist and M.~Zaldarriaga,
  Mon.\ Not.\ Roy.\ Astron.\ Soc.\  {\bf 377}, 1043 (2007)

\bibitem{Zahn:2006sg} 
  O.~Zahn, A.~Lidz, M.~McQuinn, S.~Dutta, L.~Hernquist, M.~Zaldarriaga and S.~R.~Furlanetto,
  Astrophys.\ J.\  {\bf 654}, 12 (2006)
  doi:10.1086/509597
  [astro-ph/0604177].


\bibitem{Cohn:2006ge} 
  J.~D.~Cohn and T.~C.~Chang,
  Mon.\ Not.\ Roy.\ Astron.\ Soc.\  {\bf 374}, 72 (2007)
  doi:10.1111/j.1365-2966.2006.11092.x
  [astro-ph/0603438].

\bibitem{Feng:2012mab} 
  Y.~Feng, R.~A.~C.~Croft, T.~Di Matteo and N.~Khandai,
  Mon.\ Not.\ Roy.\ Astron.\ Soc.\  {\bf 429}, 1554 (2013)
  doi:10.1093/mnras/sts447
  [arXiv:1208.6544 [astro-ph.CO]].

  
 \bibitem{Aseem:2014} 
  A.~Paranjape and T.~R.~Choudhury,
  Mon.\ Not.\ Roy.\ Astron.\ Soc.\  {\bf 442}, no. 2, 1470 (2014)
  
\bibitem{Lin:2015bcw} 
  Y.~Lin, S.~P.~Oh, S.~R.~Furlanetto and P.~M.~Sutter,
  Mon.\ Not.\ Roy.\ Astron.\ Soc.\  {\bf 461}, no. 3, 3361 (2016)

\bibitem{Giri:2017nty} 
  S.~K.~Giri, G.~Mellema, K.~L.~Dixon and I.~T.~Iliev,
  arXiv:1706.00665 [astro-ph.CO].  

  
\bibitem{Majumdar:2011uy} 
  S.~Majumdar, T.~R.~Choudhury and S.~Bharadwaj,
  Mon.\ Not.\ Roy.\ Astron.\ Soc.\  {\bf 426}, 3178 (2012)
  doi:10.1111/j.1365-2966.2012.21914.x
  [arXiv:1111.6354 [astro-ph.CO]].
  
\bibitem{Zaroubi}
 Zaroubi, Saleem, 2013, Astrophysics and Space Science Library, 396, arxiv: 1206.0267
   
\bibitem{Bowman:2018} 
 Bowman, Judd D. ~Rogers, Alan E. E.
~ Monsalve, Raul A.
 ~Mozdzen, Thomas J.
~ Mahesh, Nivedita
''Nature {\bf 555}, 67 (2018)


\bibitem{Carilli:2015yta} 
  C.~Carilli,
  PoS AASKA {\bf 14}, 171 (2015).
  



\bibitem{Ali:2005md} 
  S.~S.~Ali, S.~Bharadwaj and S.~K.~Pandey,
  Mon.\ Not.\ Roy.\ Astron.\ Soc.\  {\bf 366}, 213 (2006)

\bibitem{Yoshiura:2014ria} 
  S.~Yoshiura, H.~Shimabukuro, K.~Takahashi, R.~Momose, H.~Nakanishi and H.~Imai,  Mon.\ Not.\ Roy.\ Astron.\ Soc.\  {\bf 451}, no. 1, 266 (2015)
  
\bibitem{Shimabukuro:2015iqa} 
  H.~Shimabukuro, S.~Yoshiura, K.~Takahashi, S.~Yokoyama and K.~Ichiki,
  Mon.\ Not.\ Roy.\ Astron.\ Soc.\  {\bf 458}, no. 3, 3003 (2016)

\bibitem{Shimabukuro:2016viy} 
  H.~Shimabukuro, S.~Yoshiura, K.~Takahashi, S.~Yokoyama and K.~Ichiki,
  Mon.\ Not.\ Roy.\ Astron.\ Soc.\  {\bf 468}, no. 2, 1542 (2017)
  
\bibitem{Majumdar:2017tdm} 
  S.~Majumdar, J.~R.~Pritchard, R.~Mondal, C.~A.~Watkinson, S.~Bharadwaj and G.~Mellema,
  arXiv:1708.08458 [astro-ph.CO].


  
\bibitem{McMullen:1997} P. McMullen, 
  Rend. Circ. Palermo, {\bf 50} 259 (1997).

\bibitem{Alesker:1999}
  S. Alesker, 
  {{\em Geom. Dedicata} {\bf 74} 241-248 (1999)}.

\bibitem{Beisbart:2002}
  C. Beisbart, R. Dahlke, K. Mecke and H. Wagner, 
  Vol. 600 of Lecture Notes in Physics pp. 249-271

  \bibitem{Hug:2008}
  D. Hug, R. Schneider and R. Schuster, {\em The space of isometry covariant tensor valuations,} 
  {{\em Math. J.} {\bf 19} 137-158 (2008)}.

  
\bibitem{Schroder2D:2009}
  G.E. Schroder-Turk, S. Kapfer, B. Breidenbach, C. Beisbart, and K. Mecke,
  {{\em J. Microsc.} {\bf 238} 57 (2010)}.

\bibitem{Schroder3D:2013}
  G.E. Schroder-Turk, W. Mickel, S.C. Kapfer, F.M. Schaller, B. Breidenbach, D. Hyg, and K. Meche,
  {{\em New J. Phys.} {\bf 15} 083028 (2013)}.




    
\bibitem{Vidhya:2016} V.~Ganesan and P. Chingangbam, 
  {{\em JCAP} {\bf 1706} 023 (2017)} 

\bibitem{Chingangbam:2017} P.~Chingangbam, K.~P.~Yogendran, Joby P.~K.~, V.~Ganesan and Stephen Appleby, Changbom Park, 
  {{\em JCAP} {\bf 1712} 023 (2017)} 

\bibitem{Appleby:2017uvb} 
  S.~Appleby, P.~Chingangbam, C.~Park, S.~E.~Hong, J.~Kim and V.~Ganesan,
  arXiv:1712.07466 [astro-ph.CO].
  

\bibitem{Tomita:1986} H.~Tomita, Progr.~Theor.~Phys.~{\bf 76}, 952 (1986).

\bibitem{Gott:1990} J.~R.~Gott, C.~Park, R.~Juzkiewicz, W.~E.~Bies,
  F.~R.~Bouchet and A.~Stebbins, Astrophys.~J. {\bf 352}, 1  (1990). 

\bibitem{Mecke:1994}
  K. R. Mecke, T. Buchert, and H. Wagner, 
  {{\em Astron. Astrophys.} {\bf 288} 697 (1994)} 
   
\bibitem{Schmalzing:1997}
  J. Schmalzing and T. Buchert, 
  {{\em Astrophys. J.} {\bf 482} L1-L4 (1997)}

\bibitem{Schmalzing:1998}
  J. Schmalzing and K. M. Gorski, 
  {{\em Mon. Not. Roy. Astron. Soc.} {\bf 297} 355 (1998)} 
    
\bibitem{Winitzki:1998}
  S.~Winitzki and A.~Kosowsky,
  New Astron.\  {\bf 3}, 75 (1998)
  
\bibitem{Matsubara:2003yt}
  T. Matsubara, 
  {{\em Astrophys. J.} {\bf 584} 1 (2003)}
  

\bibitem{COBE_NG:2000}
  D. Novikov, J. Schmalzing and V. F. Mukhanov, {\em On Non-Gaussianity in the Cosmic Microwave Background}, 
    {\em Astron. Astrophys.} {\bf 364} 17 (2000) 

\bibitem{Park:2002} C.-G. Park anf C. Park, {\em JKAS} {\bf 35} 67 (2002)
  
\bibitem{Komatsu:2011} E.~Komatsu, {\em et. al.},  ApJS, 192, 18 (2011)

\bibitem{Ade:2015ava} 
  P.~A.~R.~Ade {\it et al.} [Planck Collaboration],  {\em Planck 2015 results. XVII. Constraints on primordial non-Gaussianity}, {\em Astron. and Astrophys.} {\bf 594} A17 (2016)


\bibitem{Vidhya:2014} 
Vidhya G., P. Chingangbam, K.~P.~Yogendra and  C.~Park,  JCAP, 02 028 (2015) 

\bibitem{Chingangbam:2012wp} 
  P.~Chingangbam and C.~Park,
  JCAP {\bf 1302}, 031 (2013)

\bibitem{Buchert:2017uup} 
  T.~Buchert, M.~J.~France and F.~Steiner,
Class. Quantum Grav. 34, 094002 (2017)

\bibitem{Lee:2007dt} 
  K.~G.~Lee, R.~Cen, J.~R.~Gott, III and H.~Trac,
  Astrophys.\ J.\  {\bf 675}, 8 (2008)

\bibitem{Friedrich:2010nq} 
  M.~M.~Friedrich, G.~Mellema, M.~A.~Alvarez, P.~R.~Shapiro and I.~T.~Iliev,
  Mon.\ Not.\ Roy.\ Astron.\ Soc.\  {\bf 413}, 1353 (2011)
  
\bibitem{Ahn:2010hg} 
  S.~E.~Hong, K.~Ahn, C.~Park, J.~Kim, I.~T.~Iliev and G.~Mellema,
  J.\ Korean Astron.\ Soc.\  {\bf 47}, no. 2, 49 (2014)

\bibitem{Wang:2015dna} 
  Y.~Wang, C.~Park, Y.~Xu, X.~Chen and J.~Kim,
  Astrophys.\ J.\  {\bf 814}, no. 1, 6 (2015)

\bibitem{Gleser:2006su} 
  L.~Gleser, A.~Nusser, B.~Ciardi and V.~Desjacques,
  Mon.\ Not.\ Roy.\ Astron.\ Soc.\  {\bf 370}, 1329 (2006)
  doi:10.1111/j.1365-2966.2006.10556.x
  [astro-ph/0602616]. 
 \bibitem{Yoshiura:2015} S. Yoshiura, H. Shimabukuro, K. Takahashi and T. Matsubara, arxiv:1602.0235.

  
   



\bibitem{Zahn:2010yw} 
  O.~Zahn, A.~Mesinger, M.~McQuinn, H.~Trac, R.~Cen and L.~E.~Hernquist,
  Mon.\ Not.\ Roy.\ Astron.\ Soc.\  {\bf 414}, 727 (2011)

   
\bibitem{Mesinger:2010ne} 
  A.~Mesinger, S.~Furlanetto and R.~Cen,
  Mon.\ Not.\ Roy.\ Astron.\ Soc.\  {\bf 411}, 955 (2011)


\bibitem{PLANCKreio:2016} R.~Adam {\em et al}.,  {\em Planck intermediate results
XLVII. Planck constraints on reionization history}, A\&A 596, A108 (2016)

\bibitem{PLANCK:2018_cosmoparams} N.~Aghanim {\em et al.}, {\em Planck 2018 results. VI. Cosmological parameters}, Submited to A\&A, arXiv:1807.06209 [astro-ph.CO].

  
\bibitem{planck:cosmopara2015}
  Planck Collaboration: P. A. R. Ade et al., {\em Planck 2015 results. XIII. Cosmological parameters,} 
  {{\em Astron. Astrophys.} {\bf 594} A13 (2016)} 

\bibitem{Chingangbam:2012}
  P. Chingangbam, C. Park, K. P. Yogendran and R. van de Weygaert, 
  {{\em Astrophys. J.} {\bf 755} 122 (2012)} 

\bibitem{Park:2013}
  C. Park, P. Pranav, P. Chingangbam, R. van de Weygaert, B. Jones, G. Vegter, I. Kim, J. Hidding and W. A. Hellwing, 
  {{\em JKAS} {\bf 46} 125 (2013)}



\end{thebibliography}
\end{document}